\documentclass[aps,prd,superscriptaddress,10pt,showpacs,notitlepage]{revtex4}
	\usepackage{graphicx}
	\usepackage{amsmath,amssymb,mathrsfs}
\usepackage{epsf}
\usepackage{epsfig}

\usepackage{color}

\usepackage{amsmath}

\usepackage{amssymb}

\numberwithin{equation}{section}

\usepackage{subfigure}

\newcommand{\be}{\begin{equation}}
\newcommand{\ee}{\end{equation}}
\newcommand{\bea}{\begin{eqnarray}}
\newcommand{\eea}{\end{eqnarray}}

\newcommand{\bb}{\bibitem}
 
\newcommand{\eqn}{\begin{eqnarray}}
\newcommand{\eqnx}{\end{eqnarray}}

\def\abs#1{\lvert#1\rvert} 
\DeclareMathOperator{\im}{im}

\begin{document}
\title{Roper resonances and quasi-normal modes of Skyrmions}

\author{C. Adam}
\affiliation{Departamento de F\'isica de Part\'iculas, Universidad de Santiago de Compostela and Instituto Galego de F\'isica de Altas Enerxias (IGFAE) E-15782 Santiago de Compostela, Spain}
\author{M. Haberichter}
\affiliation{Institut f{\"{u}}r Physik, Universit{\"{a}}t Oldenburg,
Postfach 2503 D-26111 Oldenburg, Germany}
\affiliation{Department of Mathematics and Statistics, University of
Massachusetts, Amherst, Massachusetts 01003-4515, USA}
\author{T. Romanczukiewicz}
\affiliation{Institute of Physics,  Jagiellonian University,
Lojasiewicza 11, Krak\'{o}w, Poland}
\author{A. Wereszczynski}
\affiliation{Institute of Physics,  Jagiellonian University,
Lojasiewicza 11, Krak\'{o}w, Poland}

\begin{abstract}
Radial vibrations of charge one hedgehog Skyrmions in the full Skyrme model are analysed. We investigate how the properties of the lowest resonance modes (quasi normal modes) - their frequencies and widths - depend on the form of the potential (value of the pion mass as well as the addition of further potentials) and on the inclusion of the sextic term. Then we consider the inverse problem, where certain values for the frequencies and widths are imposed, and the field theoretic Skyrme model potential giving rise to them is reconstructed. This latter method allows to reproduce the physical Roper resonances, as well as further physical properties of nucleons, with high precision.  
\end{abstract}
\maketitle 

\section{Introduction}
The Skyrme model \cite{skyrme} is an effective field theory (EFT) which bridges the underlying fundamental theory, Quantum Chromodynamics (QCD) - well understood in the perturbative, high energy regime - with the non-perturbative low energy region, beyond a scale where confinement and hadronisation leave only color-less states as observable particles.  The natural field degrees of freedom in this regime are the lightest quasi-particles i.e., pions. The main attractiveness of the model originates from the fact that this field content is sufficient to describe, in principle, all other excitations - baryons and atomic nuclei - which emerge as {\it non-perturbative} states in such a mesonic fluid, or in the modern language, as topological solitons. 

This solitonic framework received further support from the large $N_c$ limit where it has been rigorously shown that QCD can be described by a weakly interacting theory of mesons \cite{thooft}. Moreover, the pertinent topological index of the Skyrme model has been identified with the baryon charge. Finally, after the semiclassical quantization of zero modes of the classical solutions of the Skyrme model (Skyrmions) in a given topological sector (baryon charge) one got access to fermionic excitations of this classically purely bosonic theory. This opened the way for a realistic application of the Skyrme model for the description of baryons \cite{bary}, \cite{more-baryon} (proton, neutron, $\Delta$ resonances), lighter nuclei and their excitation bands \cite{light}, \cite{lau} as well as higher nuclei, binding energies \cite{bind}-\cite{gud1} and even infinite nuclear matter which defines properties of neutron stars \cite{stars}, \cite{piette star}. 

In the baryon number one sector and with the SU(2) flavor group, there are two simple types of degrees of freedom of nucleons whose excitations can lead to new quasiparticles. First of all, an (iso)rotational excitation explains the $\Delta$ resonance. Another possibility is to excite some vibrational degrees of freedom. This leads to new states which carry the same spin and isospin quantum numbers as the nucleons, i.e., the Roper resonances. The first three reasonably well established Ropers on top of the nucleons are: $N(1440)$, $N(1710)$, $N(1880)$, which are highly short-living quasi-particles. Specifically, the first Roper $N(1440)$ has a relatively wide Breit-Wigner width $\Gamma=300$ MeV (and quite short mean life time $\tau=\hbar / \Gamma$). While for the next two we have $\Gamma=250$ MeV (although one should be aware of some uncertainties) \cite{PDG}.

The first step towards understanding the Roper resonances within the Skyrme framework is to carefully study the existence and properties of  resonance modes (quasi-normal modes) in the classical model. 

\vspace*{0.2cm} 

The most general Poincare invariant Skyrme model with a standard Hamiltonian formulation reads
\begin{equation}
\mathcal{L}= \mathcal{L}_0+\mathcal{L}_2+ \mathcal{L}_4+\mathcal{L}_6,
\end{equation}
consisting of four different terms which possesses different mathematical properties and can be related to distinct phenomenological features. The first is a potential (non-derivative) term $\mathcal{L}_0=-\mathcal{U}(U)$ where $U$ is the SU(2)-valued Skyrme field. This is the only term which is not completely fixed. It must provide a proper mass for the perturbative pionic fields but is otherwise quite arbitrary. We shall, however, always assume that $\mathcal{U}$ only depends on $\mbox{Tr}\,  U$, such that the isospin symmetry remains unbroken. Its importance has been understood quite recently in the context of binding energies \cite{bind}, \cite{Sp2}, \cite{gud1}. Secondly, we have the usual sigma-model term which is just a kinetic term for the fields
\begin{equation}
\mathcal{L}_2=-\lambda_2  L_2 = -\lambda_2 \frac{1}{2}\mbox{Tr}\; (L_\mu L^\mu),
\end{equation}
where $L_\mu \equiv U^\dagger \partial_\mu U$. Thirdly, there is a four-derivative part, the so-called Skyrme term
\begin{equation}
\mathcal{L}_4=\lambda_4  L_4=\lambda_4 \frac{1}{4}  \mbox{Tr} \; ([L_\mu , L_\nu]^2), 
\end{equation}
which was originally introduced to circumvent the Derrick argument for the non-existence of static solitons, and therefore was mandatory for the model \cite{skyrme}. Physically, this term contributes to two-particle repulsive interactions. Finally there is a six derivative term 
\be
\mathcal{L}_6=\lambda_6 L_6= -\lambda_6  (24\pi^2)^2 \mathcal{B}_\mu \mathcal{B}^\mu, \;\;\; \mathcal{B}^\mu = \frac{1}{24\pi^2} \epsilon^{\mu \nu \rho \sigma} \mbox{Tr} \; L_\nu L_\rho L_\sigma, 
\ee
where $\mathcal{B}_\mu$ is the topological (baryon) current. By construction this term, being a topological current squared, describes some coherent, multi-particle interactions. It is the leading 
term of the model in the high energy density (baryon density) limit, providing the main contribution to the mean-field equation of state in this regime \cite{pressure}. This can happen inside (bulk) of atomic nuclei but also in cold dense nuclear matter (higher pressure and/or density \cite{term}). Phenomenologically, this term can also be related to a repulsive interaction mediated by the $\omega$ pseudo-vector meson \cite{gold}, \cite{term}. 

\vspace*{0.2cm} 

Before we proceed to the main part of the paper, we briefly present the current knowledge on vibrational modes in the Skyrme model framework and the possible relation to Ropers. The simplest approach, that is the collective mode approximation, has been applied to the massless Skyrme model $\mathcal{L}_2+\mathcal{L}_4$ \cite{hajduk} and to the sextic extension $\mathcal{L}_2+\mathcal{L}_4+\mathcal{L}_6$ \cite{kaulfuss}. This has been further developed (by coupling the vibrational excitations to the rotational modes) which resulted in a derivation of the Roper states for nucleons as well as for the $\Delta$ resonance \cite{hayashi}, \cite{biedenharn}, \cite{BPS vib}.  Linear perturbation theory has also been applied \cite{kaulfuss}, \cite{zahed}, \cite{nappi}, where the main finding was the nonexistence of an oscillating mode unless a very heavy pion is considered \cite{nappi}. This result was obtained for the usual Skyrme potential and does not have to be true for other potentials \cite{piette}. In any case, it is expected that Ropers should rather be described by resonance modes \cite{zahed}, \cite{piette}. Such a resonance has, in fact, been found in the potential-less Skyrme model  $\mathcal{L}_2+\mathcal{L}_4$ \cite{bizon}, however, no relation to the Ropers or influence of the pion mass has been discussed. 

\vspace*{0.2cm} 

It is the main aim of the present work to fill the gap and carefully analyse the properties of the first (three) lowest resonance modes in the full Skyrme model. Especially, we want to perform our investigation for two types of generalisations of the Skyrme model which are known to cure the classical binding energy problem of the original proposal. 
\begin{enumerate}
\item First of all, we want to systematically study how a specific form of the potential influences the properties of resonance modes (frequency and width). This will be done for potentials considered by Piette et al. \cite{piette} and for an unbinding potential i.e., a potential which is known  to reduce the unphysical large binding energies of Skyrmions. Here we will also study the impact of the value of the pion mass. 
\item Secondly, we will investigate these questions in the Skyrme model where the sextic term is take into account. 
\item Finally, we shall study the inverse problem, starting from a Schr\"odinger equation (Sturm-Liouville problem) which leads to certain low-lying resonance modes close to the Roper resonances. We then reconstruct the corresponding Skyrme model (i.e., the potential).
\end{enumerate}
The paper is organized as follows. In section 2, the radial vibrations of the hedgehog Skyrmion are analyzed. We formulate the corresponding Sturm-Liouville problem with an algebraic effective potential (in terms of the unperturbed solution).  Section 3 is devoted to the investigation of the impact of different forms of the potential. In section 4 we take into account  the sextic term. The inverse problem (reconstruction) is considered in section 5. In the last section we summarize our results. 
\section{Radial vibrations of charge one Skyrmions}
\subsection{Hedgehog ansatz}
We consider the general Skyrme model $\mathcal{L}_{0246} = \mathcal{L}_0+\mathcal{L}_2+ \mathcal{L}_4+\mathcal{L}_6$. It is convenient to choose the following parameterization of the coupling constants 
\be
\lambda_2= \frac{f_\pi^2}{8}, \;\;\; \lambda_4=\frac{1}{8e^2}, \;\;\; \lambda_0=\frac{f_\pi^2}{8} m^2_\pi , \;\;\; \lambda_6= \frac{2}{(24)^2 e^4 f_\pi^2} \epsilon^2 .
\ee 
Then, we introduce the physical energy and length scales $\mathcal{E}=f_\pi/4e$ and $\ell =2/ef_\pi$. This leads to the following Lagrangian in Skyrme units
\be
L=\frac{f_\pi}{4e}  \int d^3 x  \left(\frac{1}{2}\mbox{Tr} L_\mu L^\mu - \frac{1}{16} \mbox{Tr} [L_\mu, L_\nu]^2-\epsilon^2 \pi^4 \mathcal{B}_\mu^2 -m^2 \mathcal{U}\right)
\ee
where $m=2m_\pi / (f_\pi e)$ is the pion mass in Skyrme units. The multiplicative factor in front of the integral provides a transition from Skyrme units to physical units.
In the usual parametrization $U=e^{i\xi \vec{n} \cdot \vec{\tau}}$, where $\xi$ is a real scalar and $\vec{n}$ the unit three component vector field. Here $\vec{\tau}$ are the Pauli matrices. Next, we assume the stereographic projection 
\be
\vec{n}=\frac{1}{1+|u|^2} \left( (u+u^*), -i(u-u^*), 1-|u|^2 \right)
\ee
where $u$ is a complex field. Then the derivative terms take the following form
\be
\frac{1}{2}\mbox{Tr} L_\mu L^\mu=\xi_\mu \xi^\mu + 4\sin^2\xi \frac{u_\mu \bar{u}^\mu}{(1+|u|^2)^2},
\ee
\be
\frac{1}{4} \mbox{Tr} [L_\mu, L_\nu]^2=16 \sin^2 \xi \left( \xi_\mu \xi^\mu \frac{u_\mu \bar{u}^\mu}{(1+|u|^2)^2} -\frac{\xi_\mu \bar{u}^\mu \;\xi_\mu u^\nu}{(1+|u|^2)^2} \right) + 16\sin^4 \xi  \frac{(u_\mu \bar{u}^\mu)^2-u_\mu^2 \bar{u}^2_\nu }{(1+|u|^2)^2},
\ee
\be
\left( \epsilon^{\mu \nu \rho \sigma} \mbox{Tr} L_\nu L_\rho L_\sigma \right)^2= \frac{\sin^4 \xi}{(1+|u|^2)^4} (\epsilon^{\mu \nu \rho \sigma} \xi_\nu u_\rho \bar{u}_\sigma)^2 .
\ee
The potential is assumed to be a function of the scalar field $\xi$ only.

We want to consider how the charge one Skyrmion reacts under a spherically symmetric perturbation. Such a static stable $B=1$ solution is given by the hedgehog ansatz
\be
\xi=\xi_0(r), \;\;\;  u=\tan \frac{\theta}{2} e^{i\phi}
\ee 
where the static profile $\xi_0$ solves the following ODE
\bea
2\left( r^2 +2\sin^2 + \frac{\epsilon^2\sin^4 \xi}{4r^2}\right)  \; \xi_0'' &=& \nonumber \\  -4r \xi_0' -2\sin 2\xi_0 \left( \xi_0'^2 -1-\frac{\sin^2 \xi_0}{r^2} \right)  +m^2 r^2 \mathcal{U}_\xi 
+\epsilon^2 \left( \frac{\sin^4 \xi_0}{r^3} \xi_0' - \frac{\sin^2 \xi_0 \sin 2\xi_0}{2r^2} \xi_0'^2 \right) .
\label{stat-ode}
\eea
We assume that the initial perturbation does not break the spherical symmetry of the solution. Hence the $S^2$ part of the solution ($u$ field) remains unchanged while the profile function is a function of the radial coordinate and time $\xi=\xi(r,t)$. This leads to the reduced Lagrangian
\be
L= \frac{f_\pi}{4e}  \int d^3 x   \left[ \xi_\mu^2 \left( 1+2\frac{\sin^2 \xi}{r^2} + \frac{\epsilon^2\sin^4 \xi}{4r^4} \right) - 2 \frac{\sin^2 \xi}{r^2} \left( 1 +  \frac{\sin^2 \xi}{2 r^2} \right)-m^2\; \mathcal{U} (\xi) \right] \label{red}
\ee
This equation is the starting point for our further analysis. 
\subsection{Linear perturbation}
In the linear perturbation, we consider small fluctuations around the static soliton, $\xi = \xi_0 + \eta (r,t)$. The resulting quadratic in $\eta$ part of the Lagrangian reads
\bea
L_\eta &=&\frac{f_\pi}{4e}  4\pi \int dr \; \eta_\mu^2 \left( r^2 +2\sin^2 \xi_0+ \frac{\epsilon^2\sin^4 \xi_0}{4r^2}\right)  + \left(2\sin 2\xi_0  + \frac{\epsilon^2}{r^2} \sin^3 \xi_0 \cos \xi_0 \right)\xi^0_\mu  \partial_\mu \eta^2 \nonumber \\
&+&\eta^2 \left( 2\cos 2\xi_0 ( (\xi_0^\mu)^2 -1) -\frac{1}{r^2} (\sin^2 2\xi_0 +2\cos 2\xi_0 \sin^2 \xi_0)+ \frac{\epsilon^2}{4r^2} (\xi_0^\mu)^2 (6\sin^2\xi_0-8\sin^4\xi_0) \right. \nonumber \\  &-& \left. \frac{m^2}{2} r^2 \mathcal{U}''\right) .
\eea
For the oscillation (resonance) spectrum we assume $\eta (t,r) = e^{ \omega t} \eta (r)$ with $\omega=\Omega +i\Gamma$. Then, we arrive at the following functions for the corresponding Sturm-Liouville problem (for details of the procedure, see appendix A of \cite{BPS-res})
\be
p(r) = s(r) = \left( r^2 +2\sin^2 \xi_0 + \frac{\epsilon^2\sin^4 \xi_0}{4r^2}\right) 
\ee
and
\bea
q(r)&=&\xi_0'' \left(-2\sin 2\xi_0 -\frac{\epsilon^2}{2r^2} \sin^2 \xi_0 \sin 2 \xi_0 \right)  
+ (\xi_0')^2 \left( -2\cos 2 \xi_0 -\frac{\epsilon^2}{2r^2} (3\sin^2\xi_0 -4\sin^4\xi_0) \right) \nonumber \\
&+& \xi_0' \; \frac{\epsilon^2}{r^3} \sin^2\xi_0 \sin 2\xi_0 + 2\cos 2\xi_0 +\frac{1}{r^2} (\sin^2 2\xi_0 +2\cos 2\xi_0 \sin^2 \xi_0) + \frac{m^2}{2} r^2 \mathcal{U}_{\xi \xi}
\eea
where $\xi_0$ is the static solution. Finally, we can find an exact (algebraic) form of  the effective potential for the Sturm-Liouville problem in the normal form. It contains two algebraic terms
\be
Q=Q_0+Q_{\mathcal{U}}
\ee
 a derivative contribution (it has been known only in a form which contained first and second derivatives of the field \cite{kaulfuss})
\be
Q_0= \frac{2}{r^2} -\frac{4\sin^2 \xi_0}{ \left( r^2 +2\sin^2 \xi_0+ \frac{\epsilon^2\sin^4 \xi_0}{4r^2}\right) ^2} \left( r^2+3\sin^2 \xi_0 + 3\frac{\sin^4 \xi_0}{r^2} + \epsilon^2 \frac{\sin^6\xi_0}{4r^4} \right)
\ee
and a potential part (which is new)
\bea
Q_{\mathcal{U}}= \frac{m^2 r^2}{2} \frac{1}{ \left( r^2 +2\sin^2 \xi_0+ \frac{\epsilon^2\sin^4 \xi_0}{4r^2}\right)^2 } \left[ 
 \left( r^2\right. \right. &+& 2\sin^2\xi_0  + \left. \frac{\epsilon^2\sin^4 \xi_0}{4r^2}\right) \mathcal{U}_{\xi \xi}  
 \nonumber \\ &-& \left. \sin2\xi_0 \left( 1+\frac{\epsilon^2}{4r^2}  \sin^2 \xi_0\right) \mathcal{U}_\xi
\right] .
\eea
The potential contribution is the leading part in the asymptotic limit close to the vacuum. Indeed, $Q_0$ tends to zero as $r \rightarrow \infty$ (and $\xi_0 \rightarrow 0$). In the no potential case $Q_\mathcal{U}\equiv 0$ there are no positive energy bound states (oscillating modes) \cite{bizon}. There are also no negative energy states, which guarantees the linear stability of the $B=1$ Skyrmion. Oscillating modes can appear if a potential is taken into account. Then, asymptotically the effective potential tends to
\be
Q_\infty \equiv \lim_{r \rightarrow \infty} Q = \frac{m^2}{2} \mathcal{U}_{\xi \xi} |_{\xi =0}
\ee
which amounts to a possible appearance of (positive energy) bound state. As a consequence, a finite number of discrete modes \cite{piette} can show up. It should be stressed again that in the charge one sector the oscillating radial mode is an {\it unwanted} phenomenon, which is not supported by the existence of a corresponding particle state. On the contrary, the Roper resonances should rather be described by resonance modes. This may give some restrictions for the parameters of the model (calibration) as well as for some qualitative properties of the potential.
\subsection{Finding resonances}
Quasinormal modes are the solutions to the Schr\"odinger (or Sturm-Liouville) equation
\begin{equation}\label{eq:shroed}
 -u_{rr}+Q(r)u=\omega^2u
\end{equation} 
satisfying purely outgoing boundary conditions 
\begin{equation}
 u(r\to\infty)\sim e^{-ikr}
\end{equation} 
for the solution to the wave equation wave $\eta(r,t)=e^{i\omega t}u(r)$. This condition cannot be satisfied for real values of $k$ (because of the continuity equation 
which is one of the hermiticity conditions), therefore $k\in \mathbb{C}$ and hence $\omega=\sqrt{k^2-m^2}\in\mathbb{C}$. Usually, the full solution of 
the Schr\"odinger equation has both outgoing and incoming parts
\begin{equation}
 u(r) \approx A_{in} e^{ikr}+A_{out} e^{-ikr}.
\end{equation} 
However for complex values of $k$ one of the exponents grows while the other decreases. This makes it very dificult to apply appropriate boundary 
condition. We rewrite the potential generated by the skyrmion as
\begin{equation}
 Q(r) = \frac{2}{r^2}+m^2+\delta Q(r),
\end{equation} 
where $\delta Q(r)$ vanishes faster than $r^{-2}$ as $r\to\infty$. In fact, for massive fields, $\delta Q$ vanishes exponentially fast, which is 
a desired property.
The solution can be rewritten as
\begin{equation}\label{eq:decomp}
 u(r)=A(r)u_1(r)+B(r)u_2(r),
\end{equation} 
where 
\begin{equation}
 u_1(r) = \left(1+\frac{i}{kr}\right)e^{ikr}\qquad\text{and}\qquad u_2(r)=\left(1-\frac{i}{kr}\right)e^{-ikr}
\end{equation} 
are the solutions of the equation (\ref{eq:shroed}) for $\delta Q=0$ representing in and out-going waves, respectively.
The decomposition (\ref{eq:decomp}) is not unique. Therefore we choose $A$  and $B$ to satisfy the following condition
\begin{equation}
 A'u_1+B'u_2=0,\qquad \text{so that} \qquad u' = Au_1'+Bu_2',
\end{equation} 
as if $A$ and $B$ were constant, which is true for $r\to\infty$. Equation (\ref{eq:shroed}) takes the following form
\begin{equation}
 -A'u_1'-B'u_2'+\delta Q(Au_1+Bu_2)=0.
\end{equation} 
The above first order equations can be written in matrix form as
\begin{equation}
 \frac{d}{dr}\begin{bmatrix}A\\B\end{bmatrix}=
 \delta Q\begin{bmatrix}
  u_1 & u_2\\-u_1'&-u_2'
 \end{bmatrix}^{-1}
 \begin{bmatrix}
  0 & 0\\u_1&u_2
 \end{bmatrix}\begin{bmatrix}A\\B\end{bmatrix}
\end{equation} 
or in more explicit form
\begin{equation}
 \frac{d}{dr}\begin{bmatrix}A\\B\end{bmatrix}=
 \frac{i\delta Q}{2k}
 \begin{bmatrix}
 -u_1u_2& -u_2^2\\
 u_1^2&u_1u_2
 \end{bmatrix}\begin{bmatrix}A\\B\end{bmatrix}.
\end{equation} 
The initial condition required for regular solutions at $r=0$ is $A(0)=B(0)$. 
Using the scalability of the linear equation we impose the condition  $A(0)=B(0)=1$. With these initial conditions the singular term in the equation 
is cancelled. \\
Moreover, the above equations are much easier to solve numerically than the original Eq. (\ref{eq:shroed}) since the coefficients $A$ and $B$ become 
constant very 
fast as $\delta Q$ vanishes. More precisely, the method works if 
\begin{equation}
 \delta Q(r)e^{2\abs{\im{k}}r}\to 0.
\end{equation} 
The resonance solution has no incoming part, therefore the second condition is $A(\infty)=0$.

\subsection{The full time evolution}
Our investigations of the oscillating and resonance modes in the linearised model should 
be verified in a numerical analysis of the full time dependent equation which follows from the reduced Lagrangian (\ref{red}).

To compute radial vibration spectra of $B=1$ Skyrmions fully numerically,  we use a finite difference leapfrog method \cite{Battye:2001}. We discretise Eq. (II.25) on a uniform grid with spatial grid size $r=[0,100]$  and temporal grid size $t=[0,1000]$. We choose $2\times10^4$ spatial grid points ($\Delta r=0.005$) and $2\times 10^6$ temporal grid points ($\Delta t=0.0005$). The spatial derivatives used are fourth order accurate and Neumann boundaries have been imposed. 

We create suitable initial conditions for the leapfrog time evolution code by uniformly squeezing static $B=1$ Skyrmion solutions. Static solutions for given mass parameter $m$ and given coupling constant $\epsilon$ have been obtained by solving Eq. (\ref{stat-ode}) with the collocation algorithm COLSYS \cite{Ascher:1981}. 
To create uniformly squeezed initial  conditions for our time evolution code, we
rescale the radial coordinate by $r \to s r$, where the parameter $s= 0.05$. For
given mass $m$, the profile function $\xi_0(r)$ for a static Skyrmion solution has
been  computed on the interval $r \in [0, 100 ]$ with $\Delta r = 0.005$. We use the
interpolation routine {\tt{interp1d}} from Python's {\tt{SciPy}} package \cite{SciPy} to
interpolate $\xi_0(r)$ on $r \in [0, 100 ]$ and to rescale $\xi_0 (r) \to \xi_0 (s r)$ . Each
squeezed initial condition  is then time-evolved over $2 \times 106$ timesteps with
$\Delta t = 0.0005$ on a spatial grid $r \in [0, 100]$ with $\Delta r = 0.005$. In each time
step, we measure at $r = 0.1$ (inside the soliton core) the deviation from the
static Skyrmion solution by recording $\xi (t,r = 0.1)$.

To find the frequency components, we compute the fast fourier transform (FFT) $\tilde{\xi}(\omega,r=0.1)$ of the recorded data using the MATLAB built-in function {\tt{fft}}. For visualisation purposes, we present our results in the form of contour plots of $\log(|\tilde{\xi}_t\left(\omega,r=0.1\right)|)/\log(10)$.

\section{The standard Skyrme model and the role of the potential}
\subsection{The pion mass potential}
\begin{figure}
\hspace*{-1.0cm}
\includegraphics[height=8.5cm]{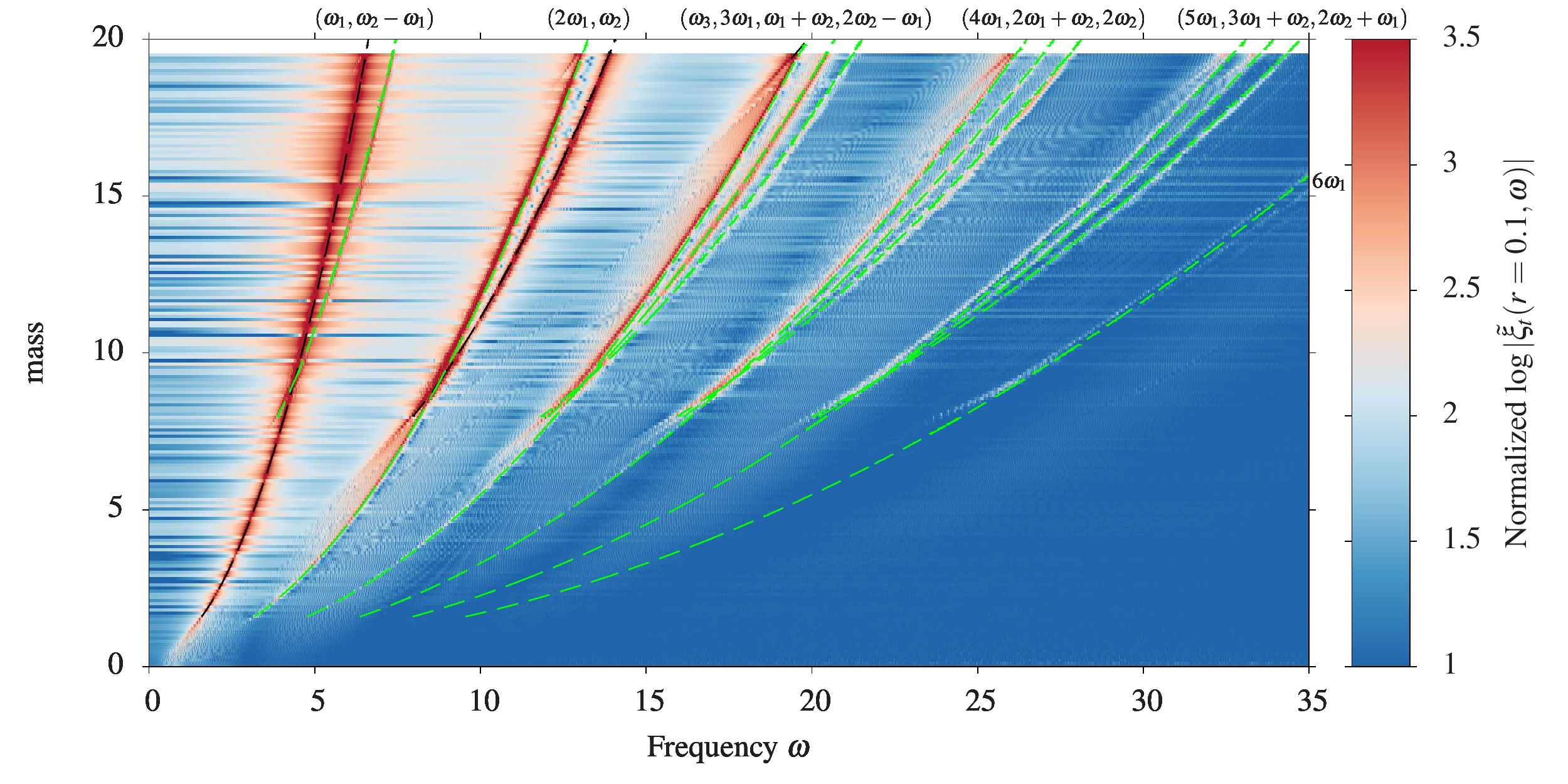}
\caption{Power spectra of the full dynamics for the usual Skyrme potential with the pion mass $m\in [0,20]$ with the first three oscillating modes (black dashed) and higher harmonics (green dashed) denoted.}
\label{scan L024}
\end{figure}

\begin{figure}
\hspace*{-1.0cm}
\includegraphics[height=9.5cm]{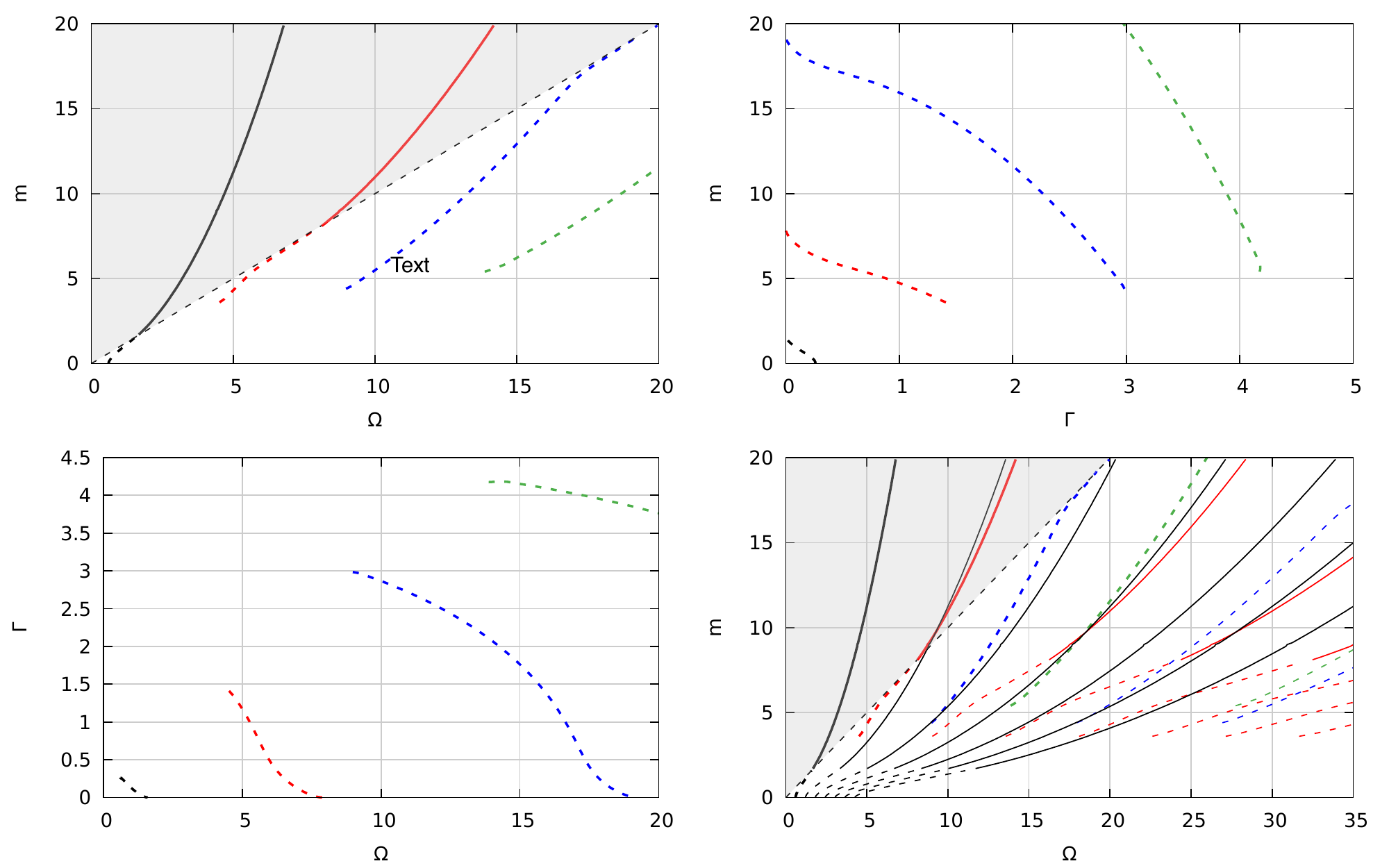}
\caption{Resonance modes for the usual Skyrme potential with the pion mass $m\in [0,20]$.}
\label{Res L024}
\end{figure}

As a first example, we consider the usual potential which provides a mass for perturbative (pionic) excitations
\be
m^2\mathcal{U}_{\pi}=2m^2(1-\cos  \xi).
\ee
As we already pointed out, this potential results in too high binding energies and, therefore, is not so suitable for a quantitative description of atomic nuclei. Here, we want to check its applicability to the Roper problem.  For reasons of generality, we vary the mass parameter from 0 to the physically much too high value 20.  Moreover, this allows us to detect resonance modes when they emerge from oscillating modes. 

In Fig. \ref{scan L024} we present a scan of the Fourier transform of  a radial perturbation of the hedgehog solution in the $\mathcal{L}_{024}$ Skyrme model with the usual potential. We vary the mass parameter from $m=0$ to $m=20$ in steps of $\Delta m=0.1$. It is clearly seen that for the (first three) oscillating modes  the linear perturbation works very well (see Fig. \ref{Res L024}) and there is no difference with the full
computation. Some higher harmonics can easily be found. In the plot it is clearly visible that the higher harmonics $n \omega_1$ exist (lower right panel). The intensity (amplitude) of higher modes obviously decreases with $n$.

The main findings are the following:
\begin{enumerate}
\item The number of oscillating modes increases with the pion mass (Fig. \ref{scan L024}). This is an obvious effect since the effective potential $Q$ forms a deeper well while $m$ grows.

\item Every resonance mode originates in a corresponding oscillating mode (Fig. \ref{Res L024}). The transition happens when, while decreasing the pion mass, the frequency of the oscillating mode meets the mass threshold. 
Indeed, when an oscillating mode passes the mass threshold line it transforms into the corresponding resonance mode. At the beginning (close to the threshold line) the mode is very narrow. It becomes wider when the mass parameter is further reduced.  

\item The critical pion masses below which the n-th oscillating mode transforms into its quasi-normal counterpart are: $m_1=1.59$, $m_2=7.99$, $m_3=19.38$ (see Table I). This means that for $m <m_1$ there are no oscillating modes at all, which sets an upper bound for the pion mass in this Skyrme model.

\item We control the first resonance mode in the full range of the pion mass (Fig. \ref{Res L024}). Its frequency starts at $\Omega=1.59$  for $m=1.59$ and gets smaller until $\Omega =0.61 $ for $m=0$ (which agrees with the original result by Bizon et. al. \cite{bizon}). Simultaneously, the width grows from basically $0$ to $\Gamma=0.26$ for $m=0$.  

\item As far as the next three resonance modes are considered, we could reduce the pion mass approximately to $m \approx 4-5$. Below this value the modes are too broad and they are beyond our accuracy. Unfortunately, this means that we are not able to reach the physical regime $m<m_1$. Nonetheless, we can use the values of the width of second, third and fourth resonance mode at the last trustable $m$ as a lower bound for the widths for physical $m$. Specifically, we found $\Gamma_2 >1.41$,  $\Gamma_3 >2.98$,  $\Gamma_4 >4.17$.  
It is clearly visible that the width of the second and higher resonance modes are much (an order of magnitude) bigger than the width of the first resonance (Fig. \ref{Res L024}). This should be contrasted with the fact that the three lowest Roper states possess very similar widths. In other words, the standard Skyrme model with the pionic mass potential fails to reproduce higher Roper resonances, at least their widths. 
\end{enumerate}
In the next two subsections we will study how these findings change if other potentials are chosen. 
\subsection{Deformed pion mass potential}

\begin{figure}
\includegraphics[height=6cm]{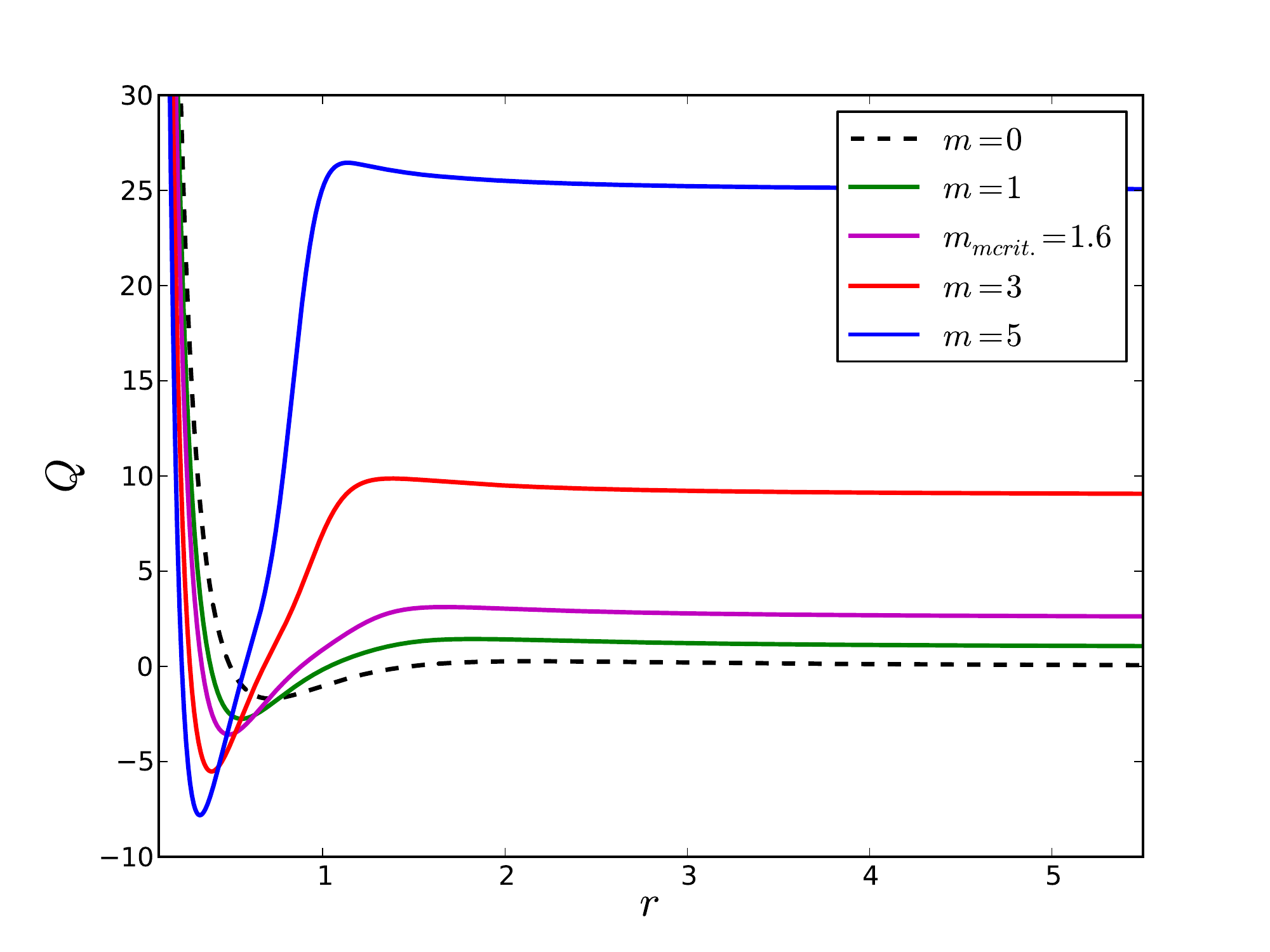}
\includegraphics[height=6cm]{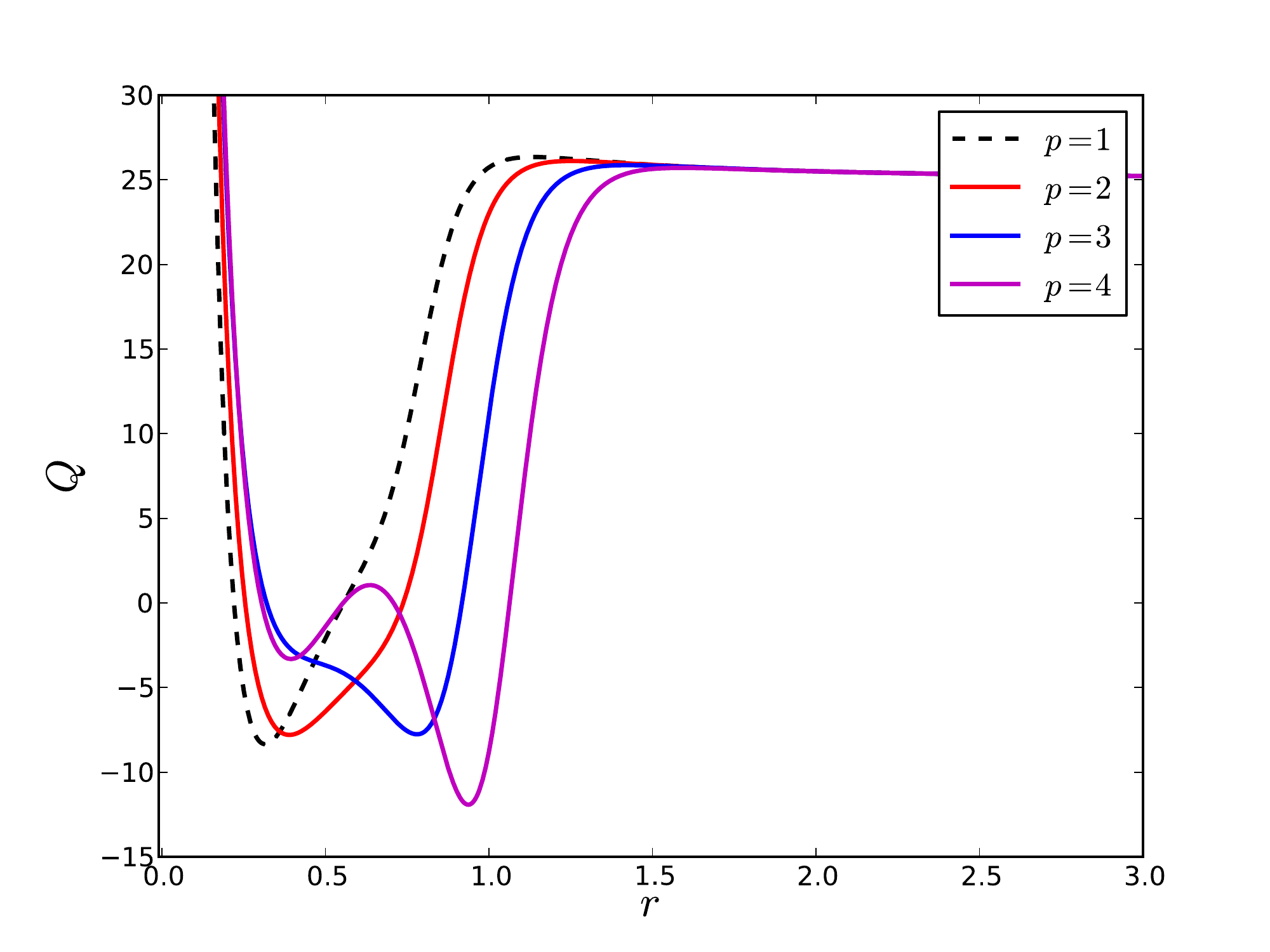}
\caption{Effective potential $Q$ for the $L_{240}$ Skyrme model. {\it Left:}  with the usual Skyrme potential and different values of the mass $m$. {\it Right:} with the potentials $\mathcal{U}^{(p)}$ and $m=5$.} \label{Q L024}
\end{figure}

As the first example of a deformation of the standard pion mass potential we refer to Lin and Piette \cite{piette} who considered a family of multi vacuum potentials ($p =1,2,3,4$)
\be
m^2\mathcal{U}^{(p)}=\frac{2m^2}{p^2}(1-\cos p \xi) ,
\ee
where $p=1$ gives the formerly considered standard potential. Qualitatively the potential $\mathcal{U}$ can influence the effective potential $Q$ (and therefore the existence and number of oscillating and resonance modes) in two simple ways. First of all, as we have already observed,  increasing the pion mass leads to a {\it higher} asymptotic value of the effective potential. As a consequence, the little potential well which exists for the massless Skyrme model $\mathcal{L}_{24}$ becomes bigger (with a higher right boundary). Hence, more oscillating modes can exist. Secondly, potentials with smaller value at the anti-vacuum ($\xi=\pi$), or in general less peaked, develop a {\it broader} well. Again, more oscillating states can emerge. Moreover, their eigen-frequencies are smaller.  All this is a direct consequence of Fig. \ref{Q L024}. 
\begin{figure}
\includegraphics[height=5cm]{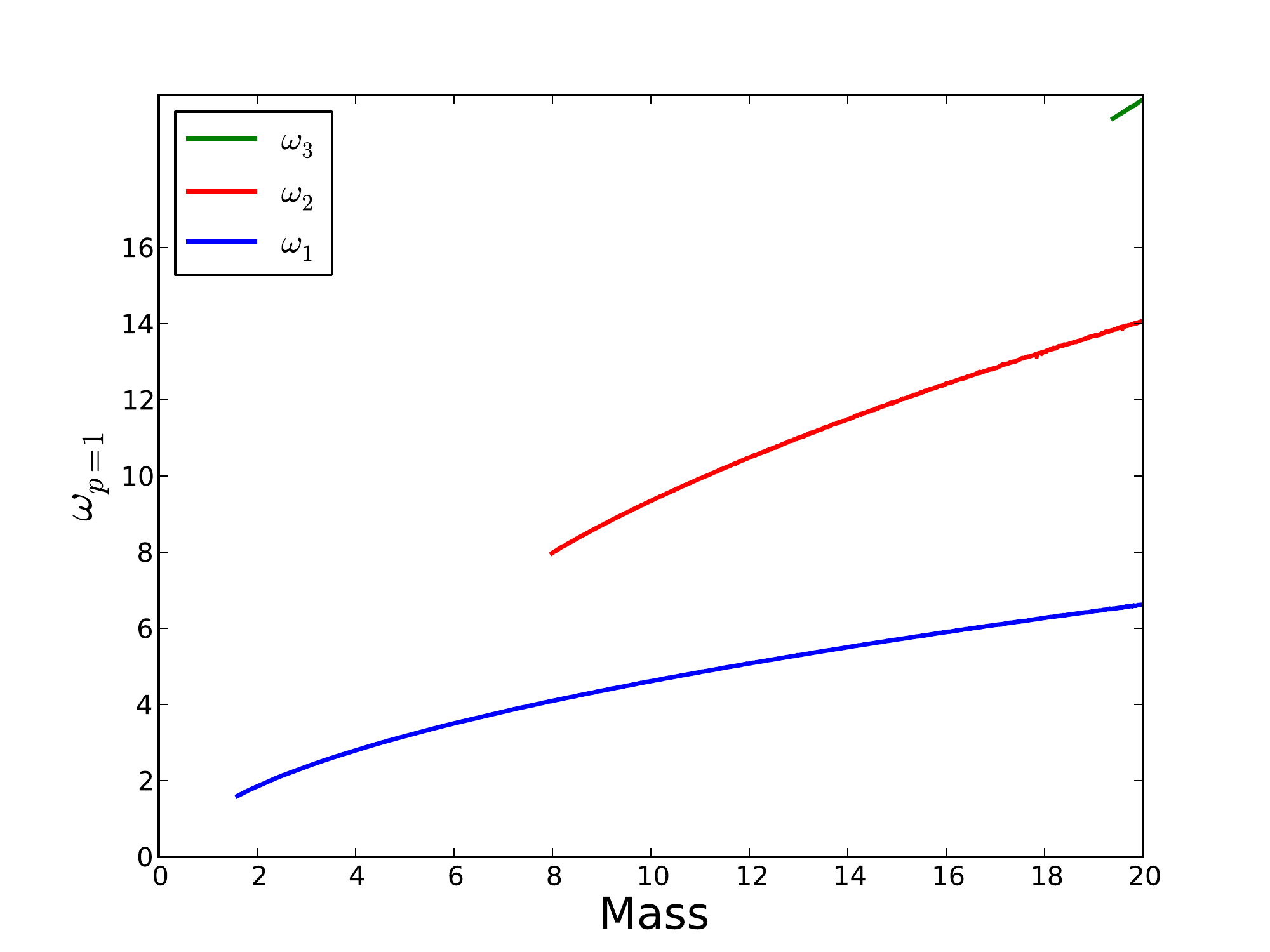}
\includegraphics[height=5cm]{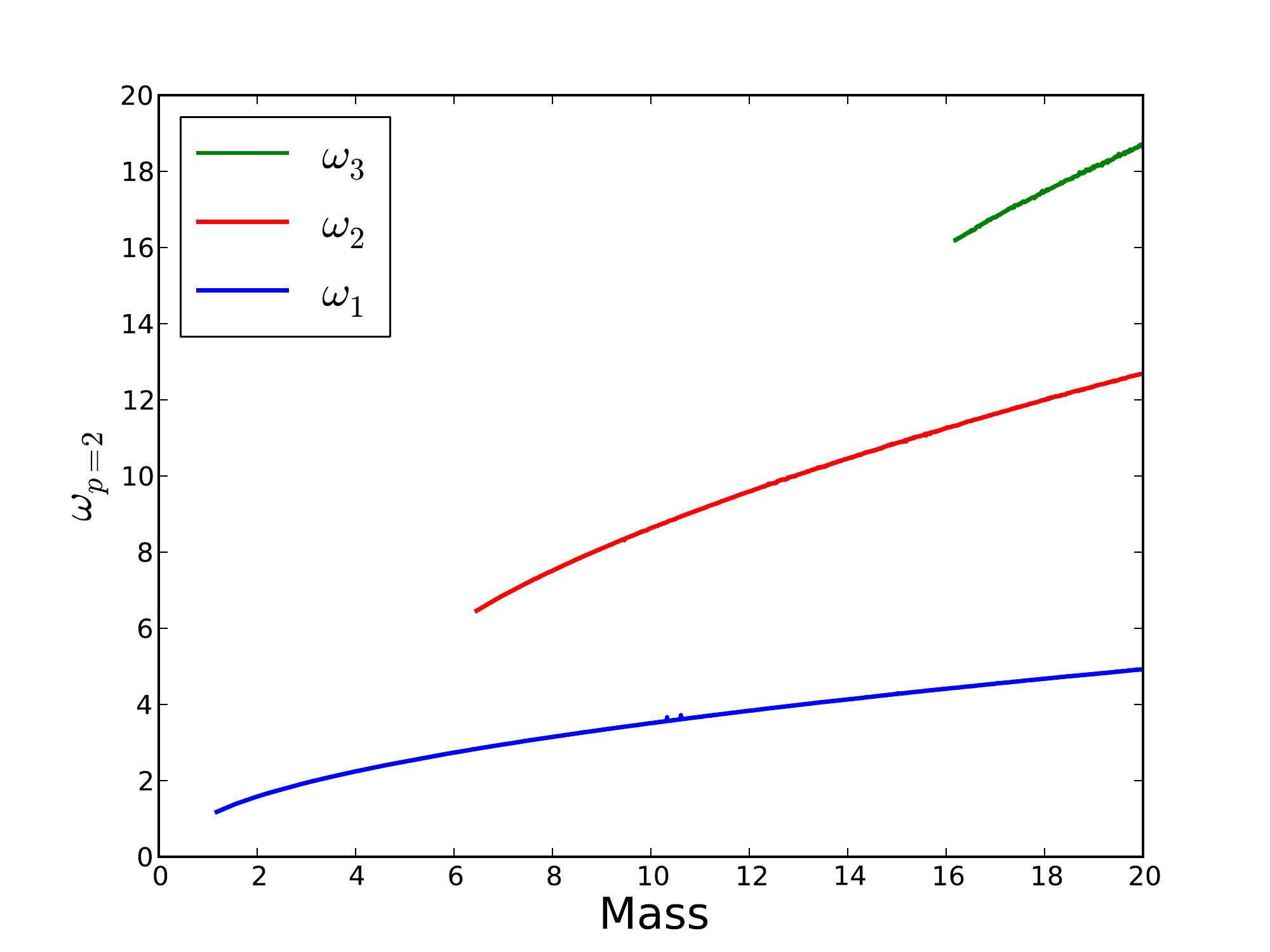}
\includegraphics[height=5cm]{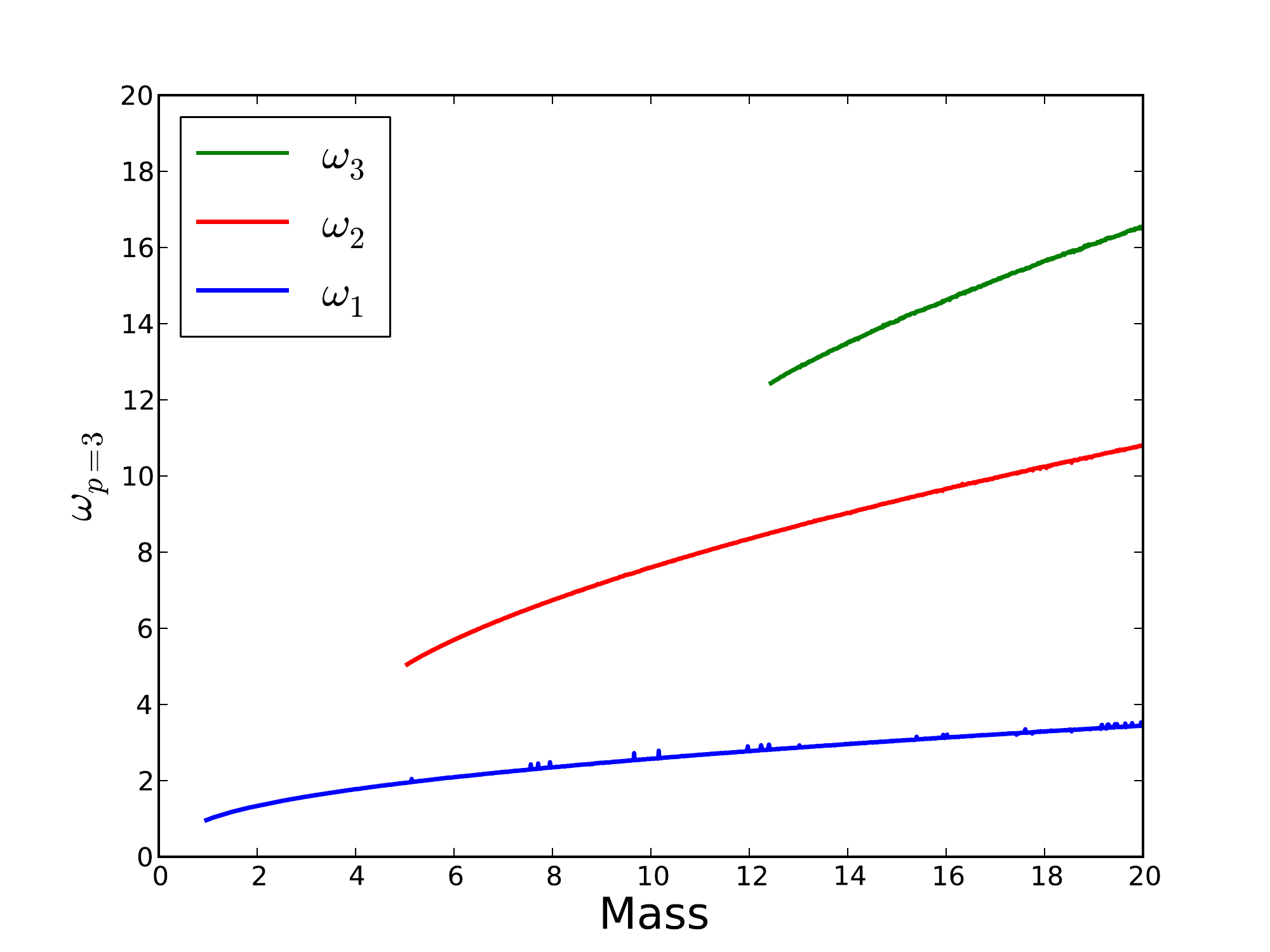}
\includegraphics[height=5cm]{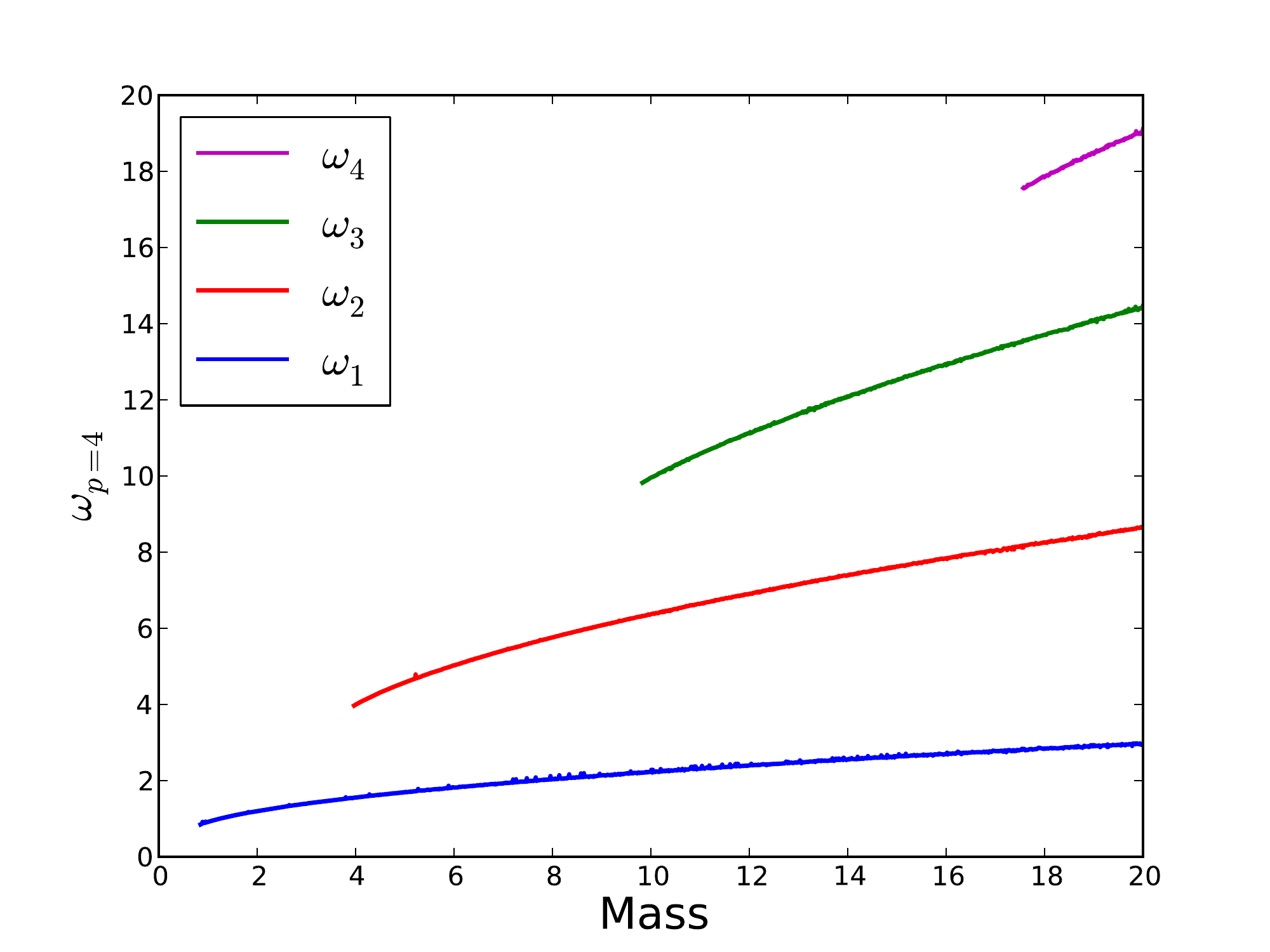}
\caption{The frequency of the first oscillating modes for $L_{024}$ Skyrme model as a function on the pion mass for Piette's potentials $p=1,2,3,4$.} \label{w L024}
\end{figure}
\begin{figure}
\hspace*{-1.0cm}
\includegraphics[height=7.5cm]{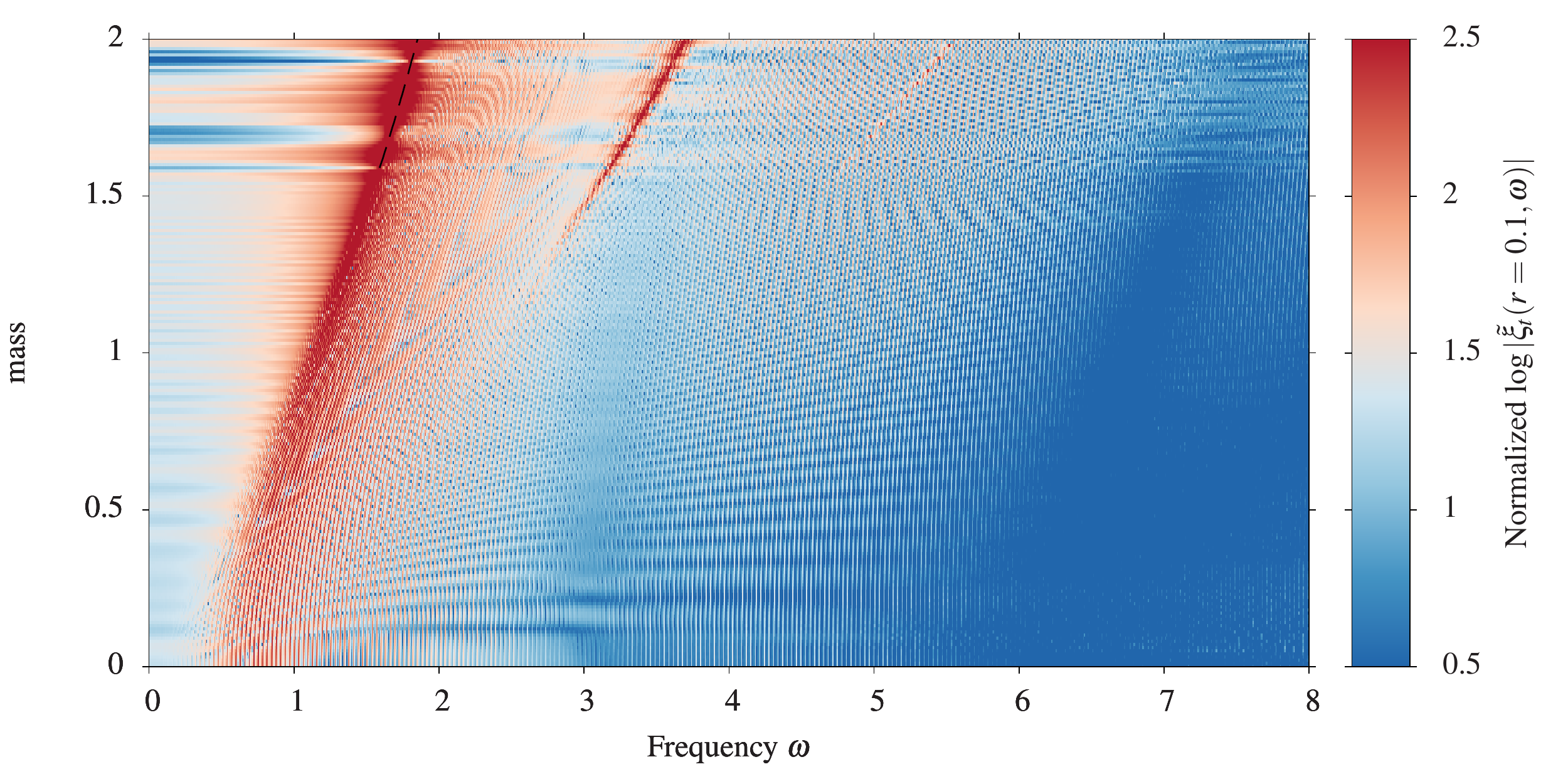}
\hspace*{-1.0cm}
\includegraphics[height=7.5cm]{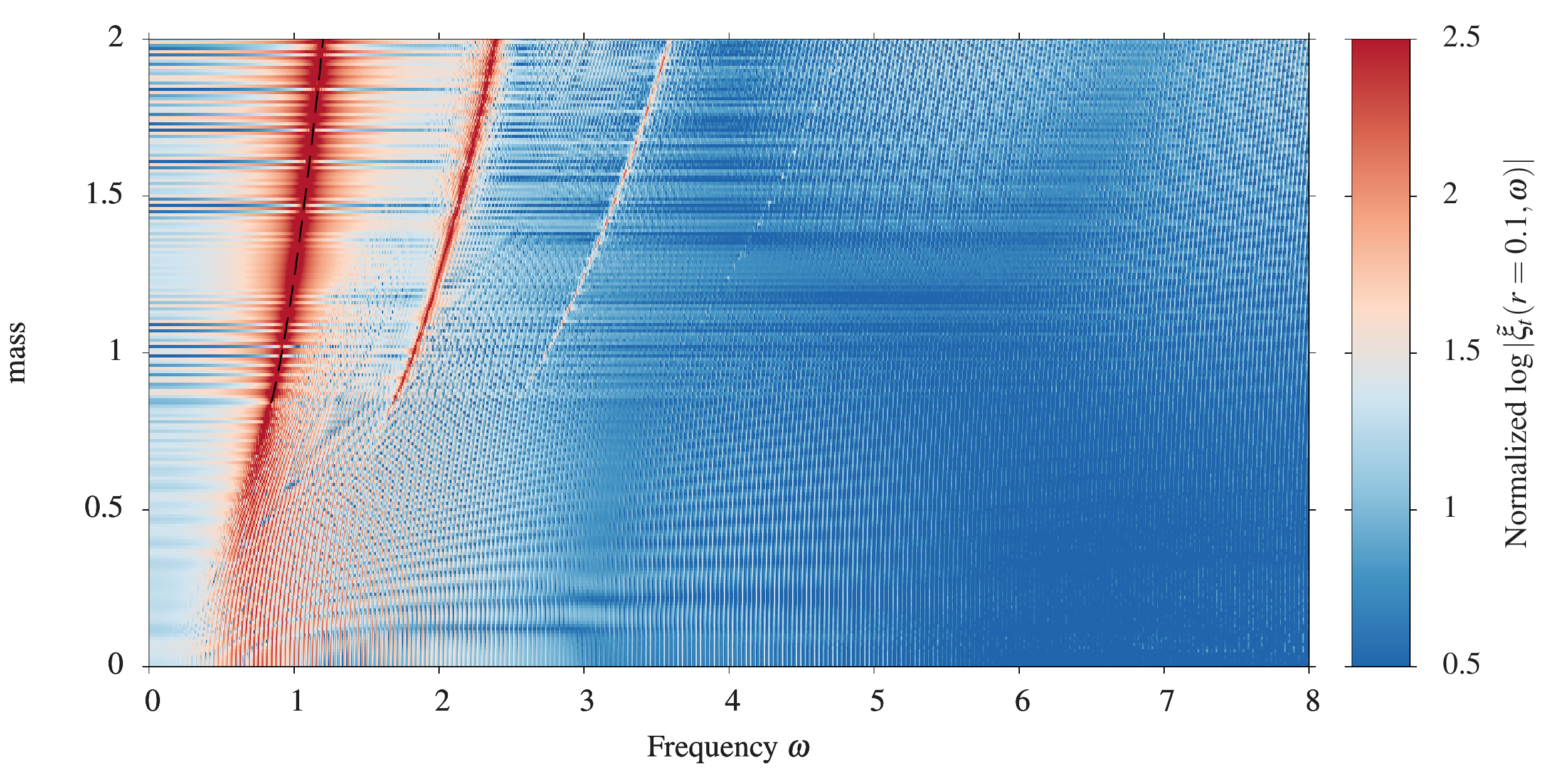}
\caption{FFT scan of the full dynamics for the $\mathcal{L}_{024}$ Skyrme model with potential $\mathcal{U}^{(1)}$ (upper) and $\mathcal{U}^{(4)}$ (lower) with the pion mass in a physical region $m\in [0,2]$. The black dashed line denotes the first oscillating mode. Here the resolution $\delta m =0.01$.}
\label{scan L024 zoom}
\end{figure}

This qualitative picture is fully confirmed by direct numerical computations - see Fig. \ref{w L024}. Quantitatively we found that the critical pion mass at which the first oscillating mode transforms into the lowest resonance mode decreases to $m=0.85$ for the $p=4$ potential. This implies that for this model the physical pion mass in Skyrme units cannot be larger than $0.85$. In Fig. \ref{scan L024 zoom} we compare the full dynamics (power spectrum) of the $p=4$ potential with the original pion mass potential. Here $m \in [0.2]$. 
\begin{table}
\begin{center}
\begin{tabular}{c|cccc}
\hline
\hline
 & $n=1$ & $n=2$ &  $n=3$&  $n=4$  \\
\hline 
\hline
$p=1 \;\;\;$  &  1.59 & 7.99 &19.38  & - \\
$p=2 \;\;\;$ &  1.18 & 6.47 & 16.22 & -  \\
$p=3 \;\;\;$ &  0.95 & 5.07 & 12.38 & -  \\
$p=4 \;\;\;$ &  0.85 & 3.98 & 9.77 & 17.56 \\
\hline
\hline
\end{tabular}  
\caption{Critical masses for the full Skyrme model with the potential $\mathcal{U}^{(p)}$. No value denotes that the critical point occurs above $m=20$.} 
\end{center}
\end{table} 

Interestingly, the higher resonances now appear much faster (for much smaller pion mass) and therefore have a chance to be much narrower and even detectable in (or close to) the physical regime, i.e., when no oscillating modes exist. Indeed, in Fig. \ref{Res p=2,4} we show the frequency and width of the lowest resonances for the $p=2$ and $p=4$ potentials. However, although the quasi-normal states are narrower, there seems to be still a significant difference between the first and higher resonances. Therefore, these potentials do not provide a reliable description of the Roper resonances, either.

\begin{figure}
\hspace*{-1.0cm}
\includegraphics[height=5.0cm]{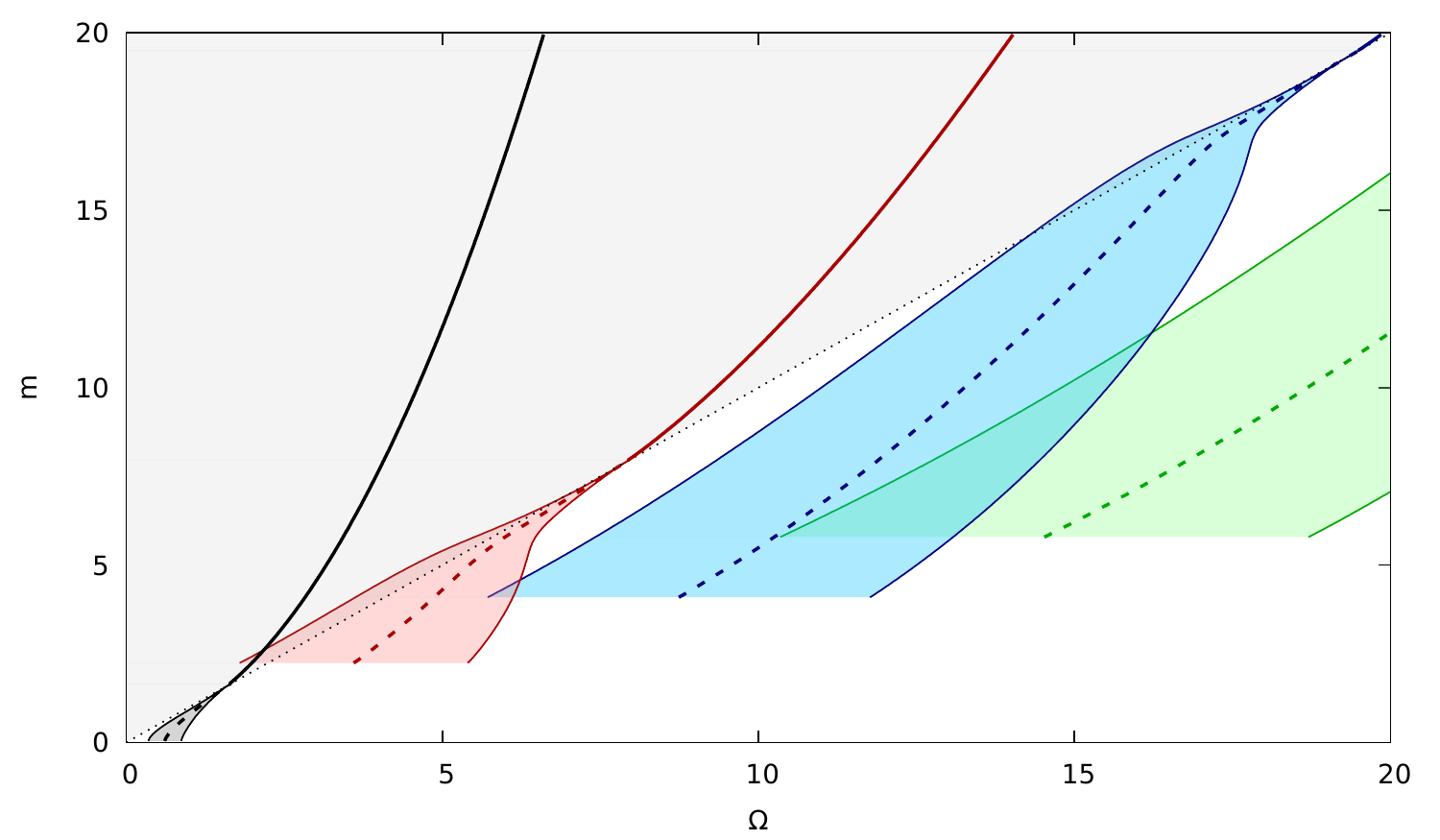}
\includegraphics[height=5.0cm]{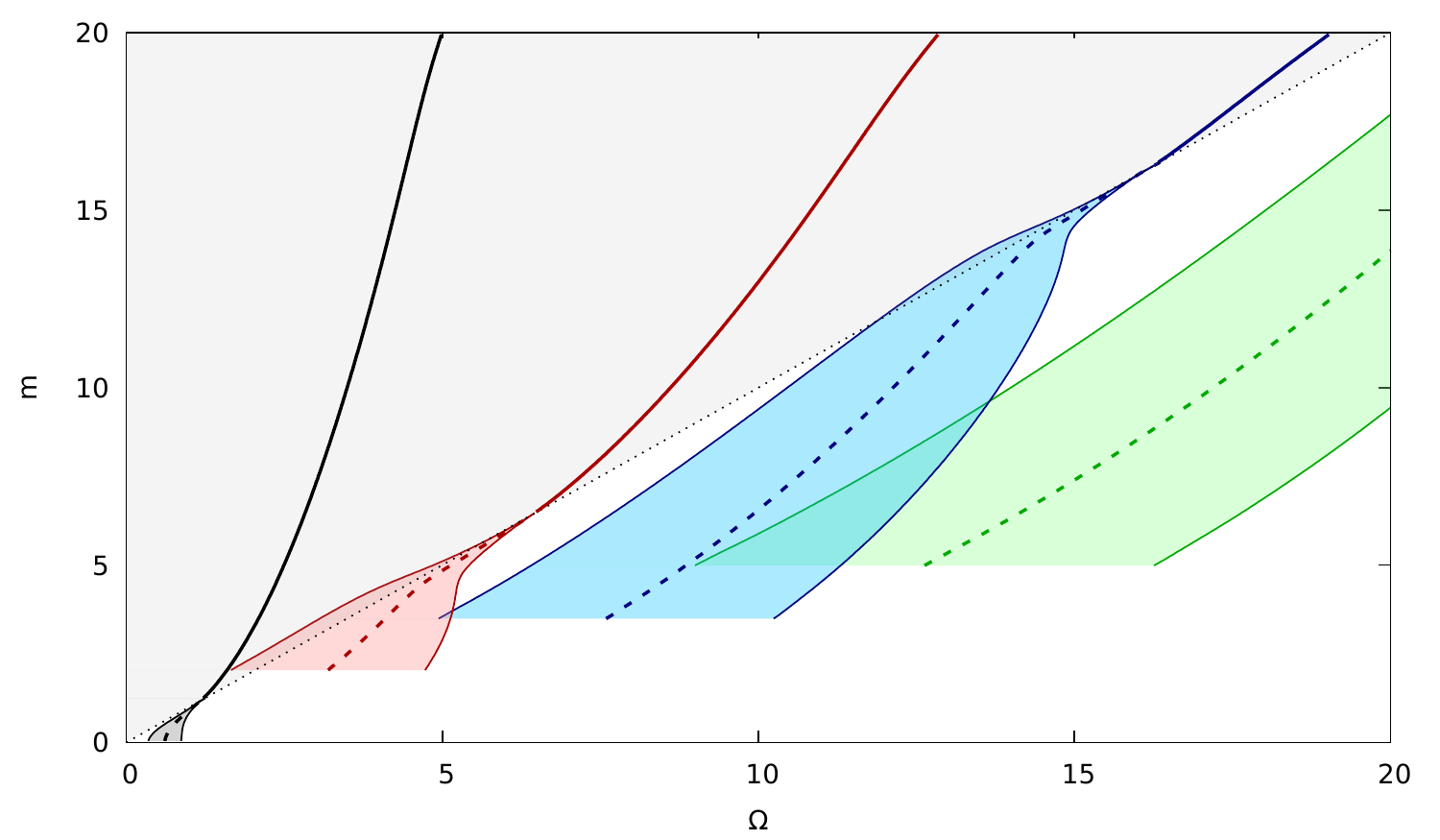}
\hspace*{-1.0cm}
\includegraphics[height=5.0cm]{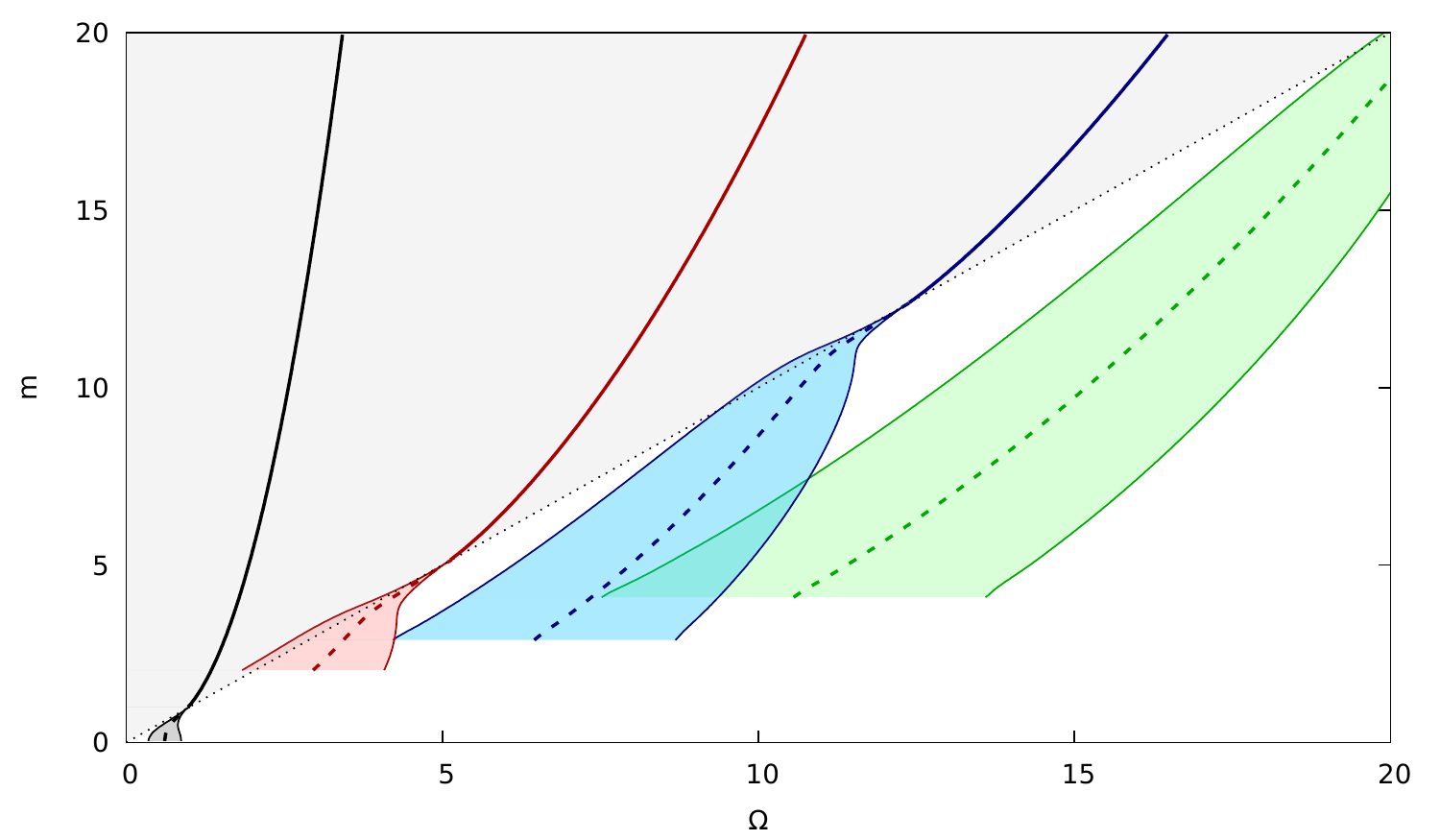}
\includegraphics[height=5.0cm]{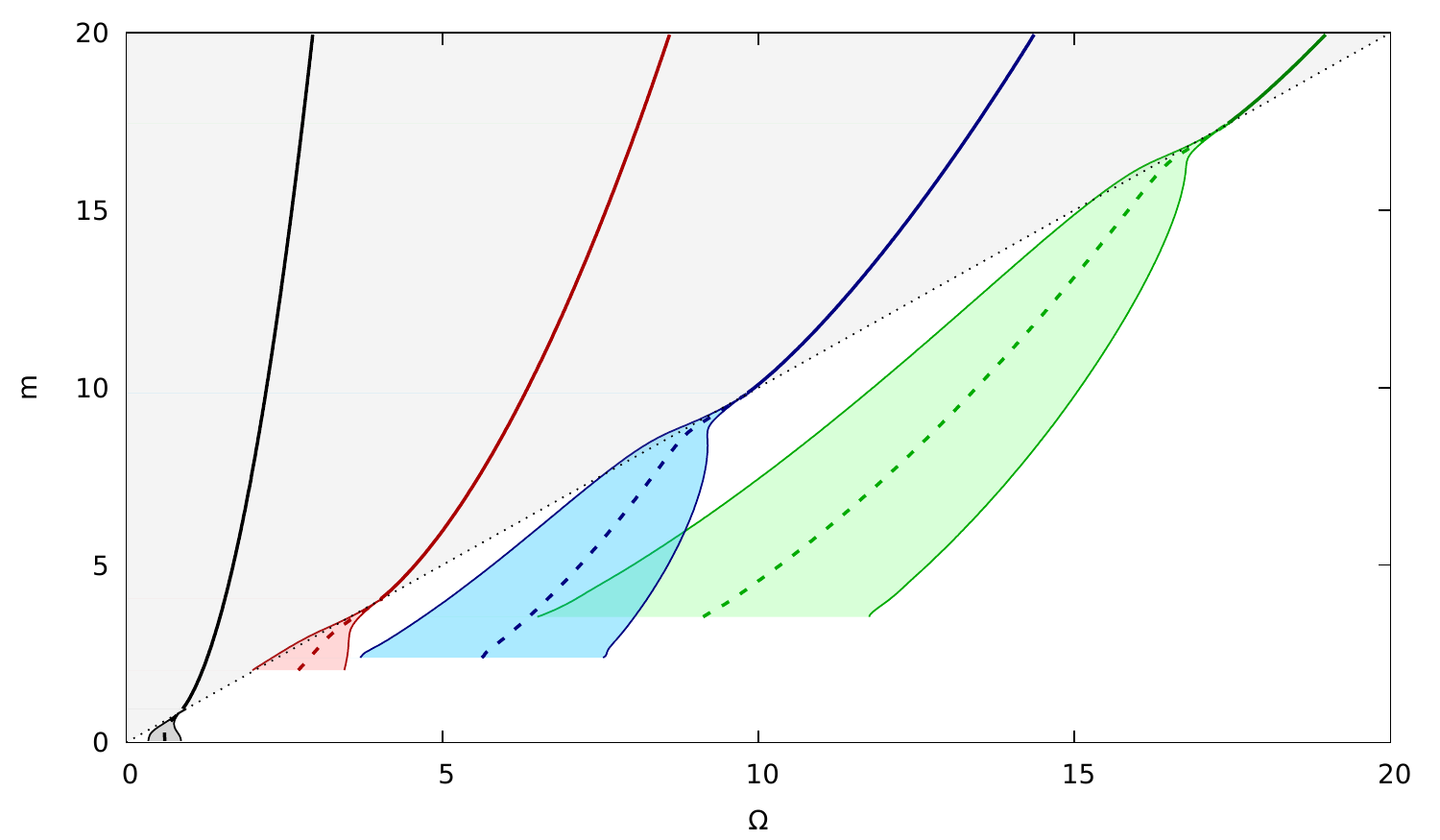}
\caption{Resonance modes for the Piette's potentials $p=1,2,3,4$ with the pion mass $m\in [0,20]$.}
\label{Res L024p}
\end{figure}

The first four oscillating/resonance state for the Piette's potentials with $p=1,2,3,4$ are plotted in Fig. \ref{Res L024p}. Clearly, our numerical resolution allows for a precise computation for the first mode. The higher resonances required bigger accuracy and, unfortunately, they are no longer under control before the first oscillating mode turns into a resonance. 

\begin{figure}
\includegraphics[height=5cm]{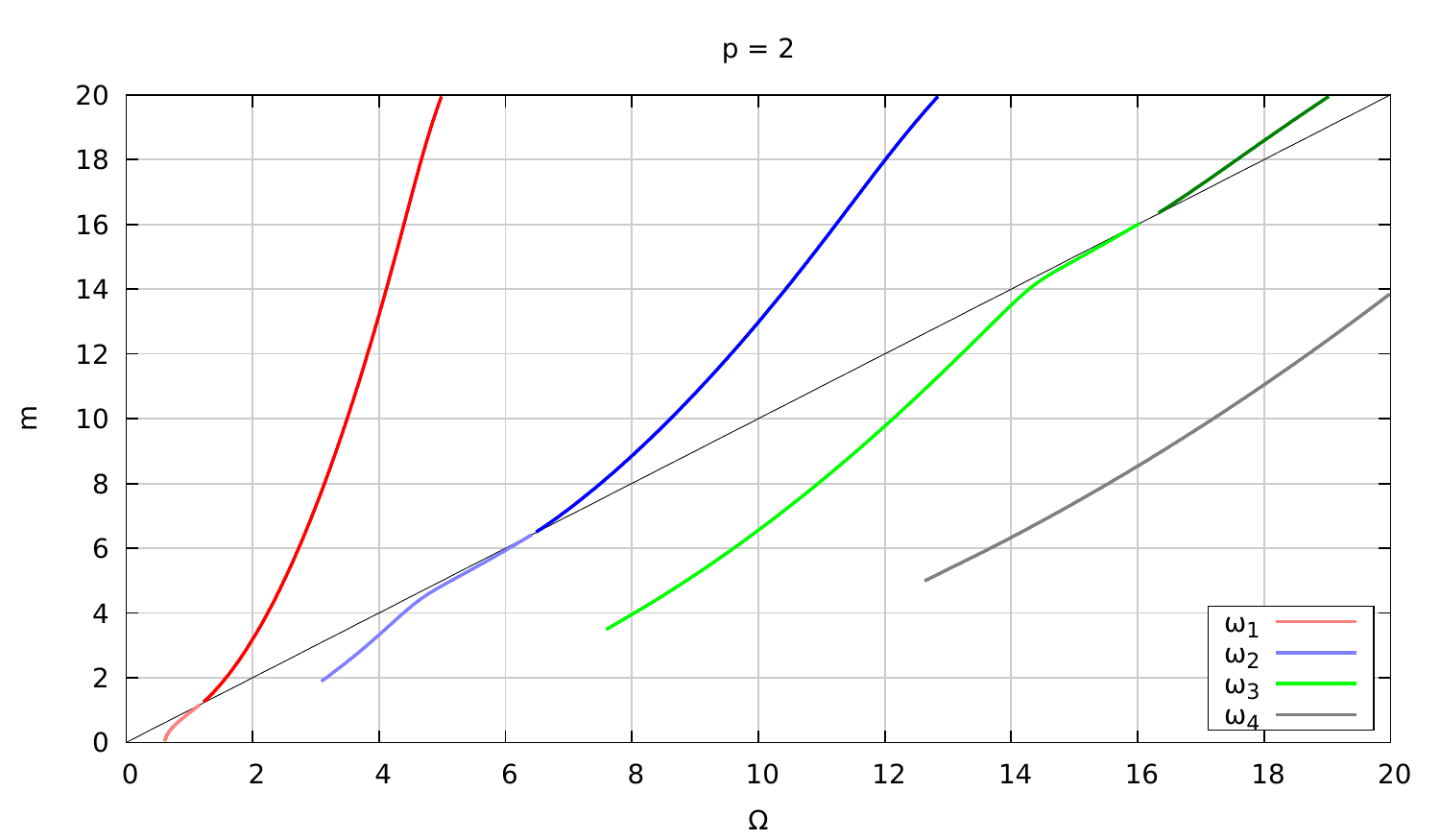}
\includegraphics[height=5cm]{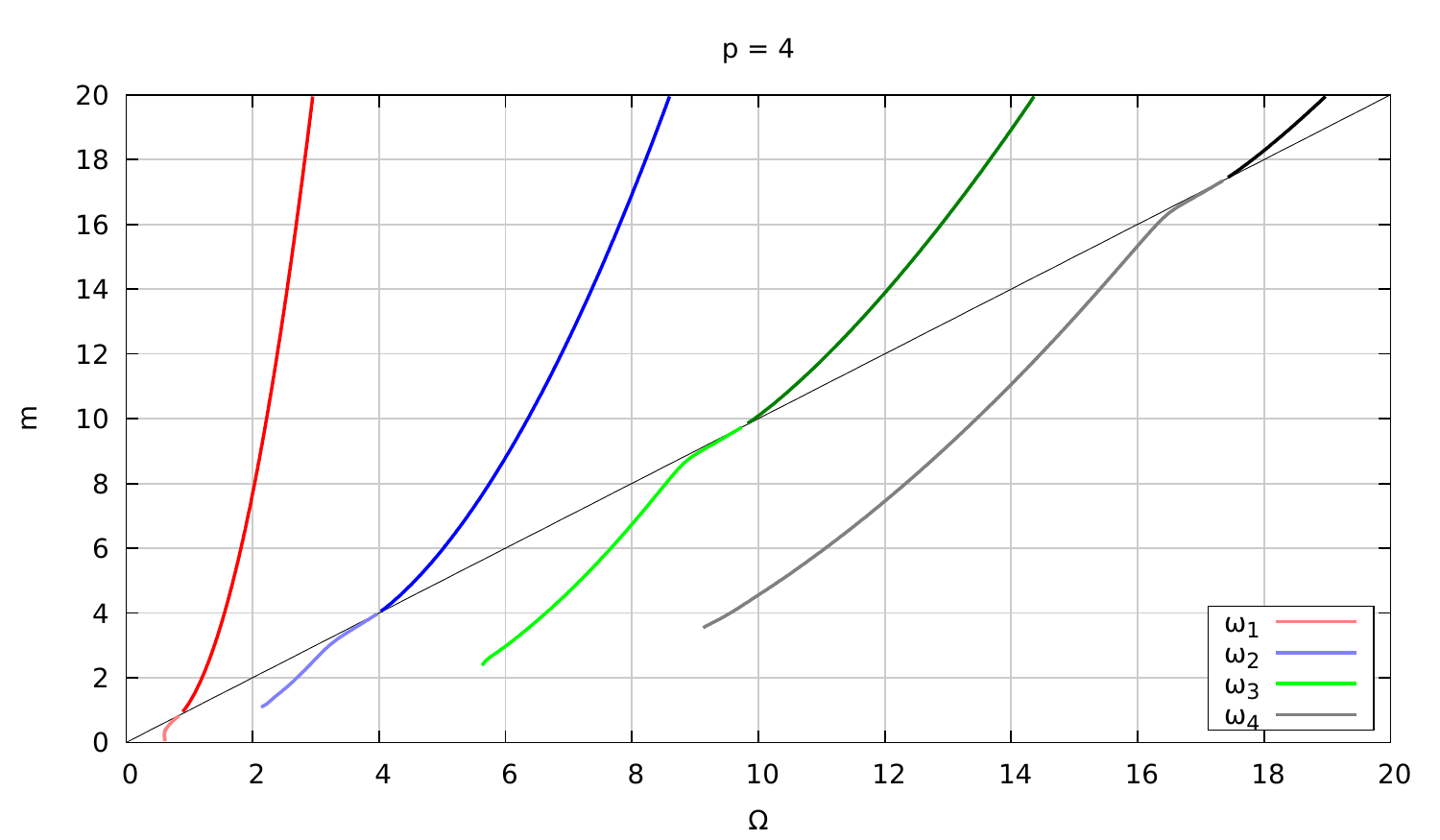}
\includegraphics[height=5cm]{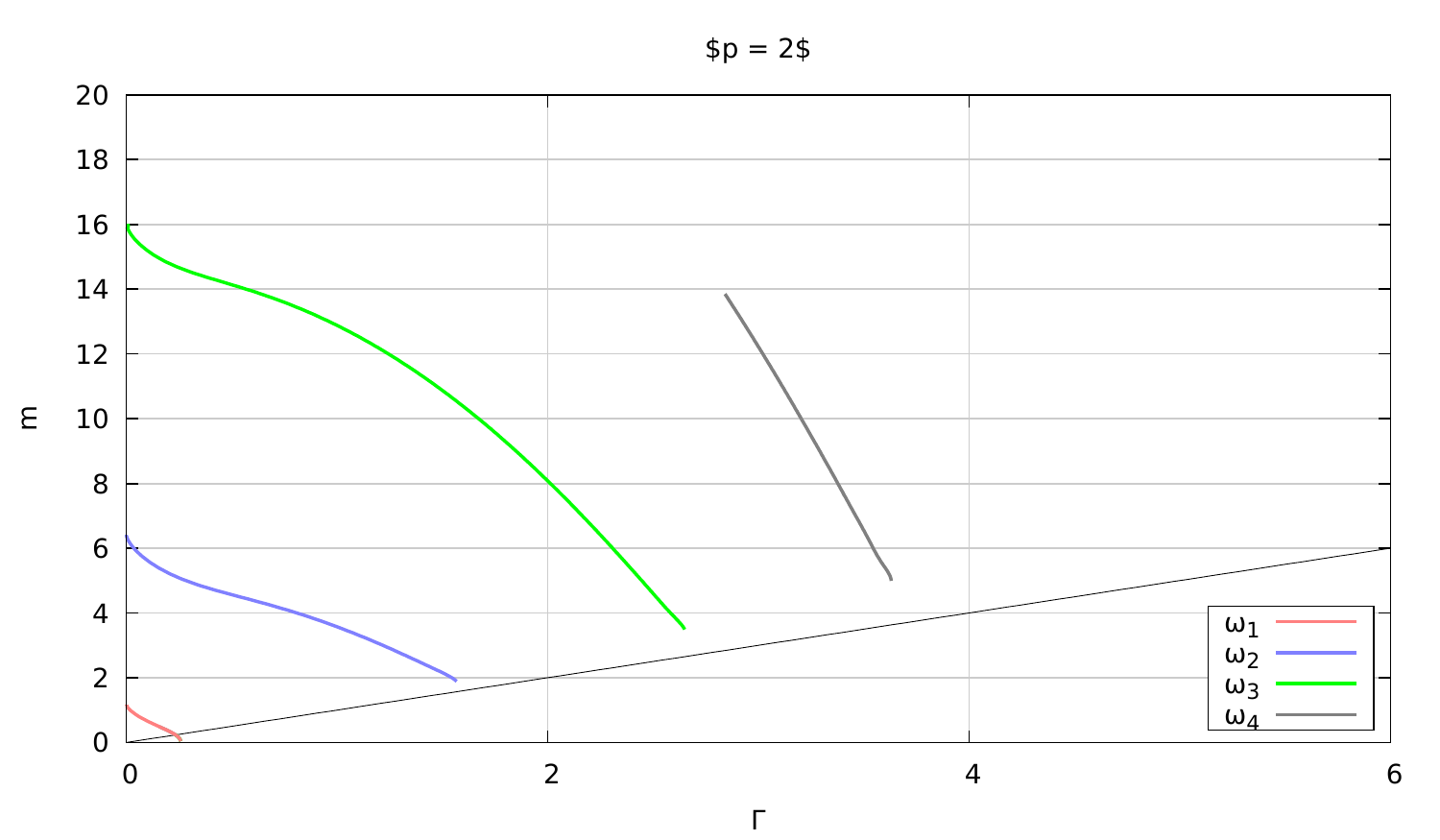}
\includegraphics[height=5cm]{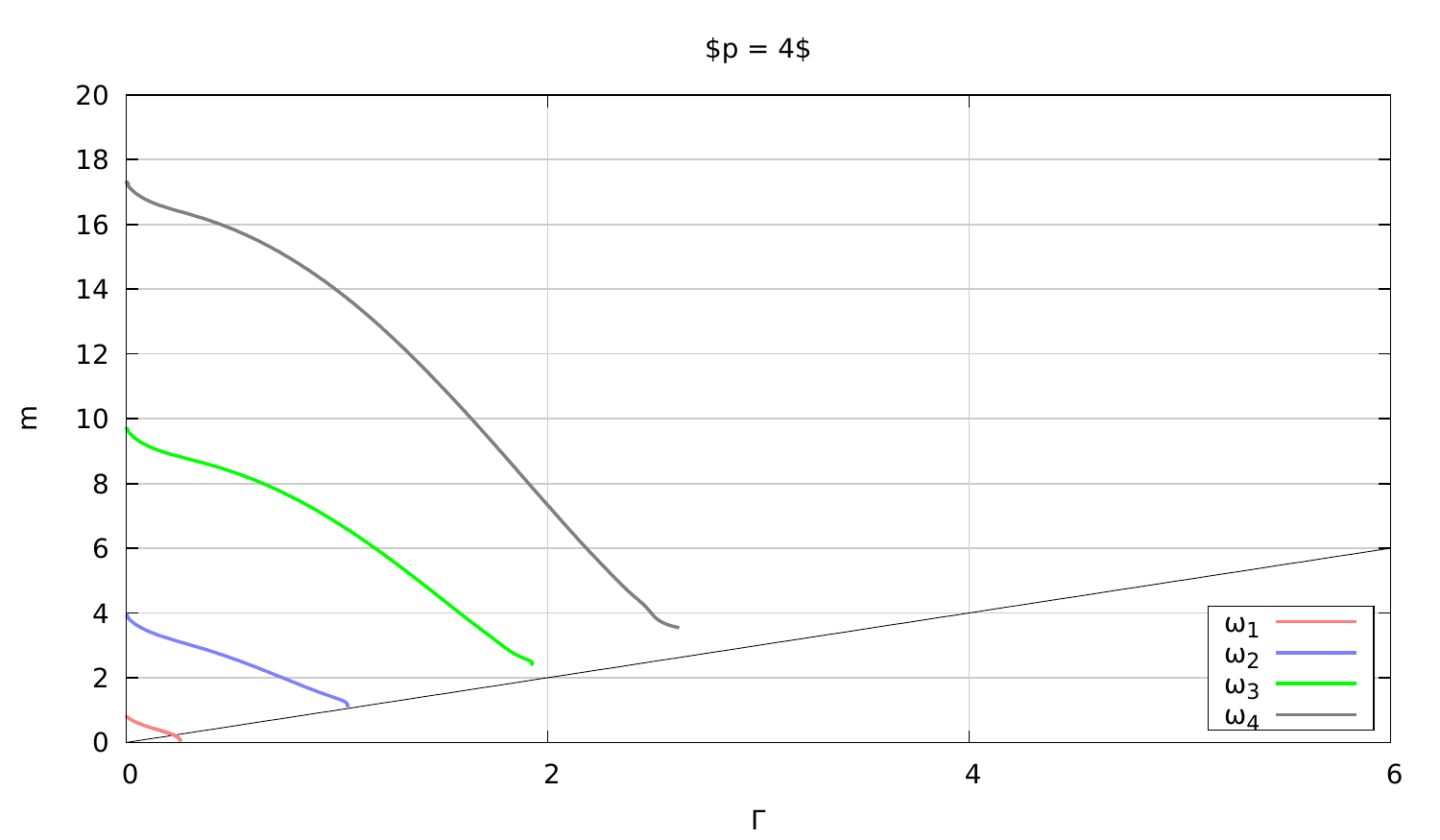}
\caption{The frequency $\Omega$ and width $\Gamma$ of the lowest resonance modes a function on the pion mass for Piette's potentials $p=2$ and $p=4$.} \label{Res p=2,4}
\end{figure}
 
\subsection{Unbinding potential}
The next type of potential is the unbinding potential considered recently  \cite{Sp2} (see also \cite{gud1})
\be
m^2\mathcal{U}_{\alpha}=2m^2(1-\cos  \xi) +\frac{\alpha}{(1-\alpha)^2} (1-\cos \xi)^4
\ee
where $\alpha <1$ is a parameter which controls how close to the unbinding regime we are. In the limit $\alpha=1$ (after some unit redefinition) we arrive at a Skyrme model which saturates a corresponding Bogomolny bound and does not have stable higher charge solitonic solutions. Therefore it describes a completely unbinding model. It has been reported that physical binding energies can be obtained if $\alpha =0.95$ and 
\be
f_\pi= 36.1 \mbox{ MeV}, \;\;\; m_\pi = 303 \mbox{ MeV}, \;\;\; e = 3.76
\ee
which correspond with $m=4.46$. Note that this is a rather big number if compared with the usual calibration scheme. 
\begin{figure}
\hspace*{-1.0cm}
\includegraphics[height=5.0cm]{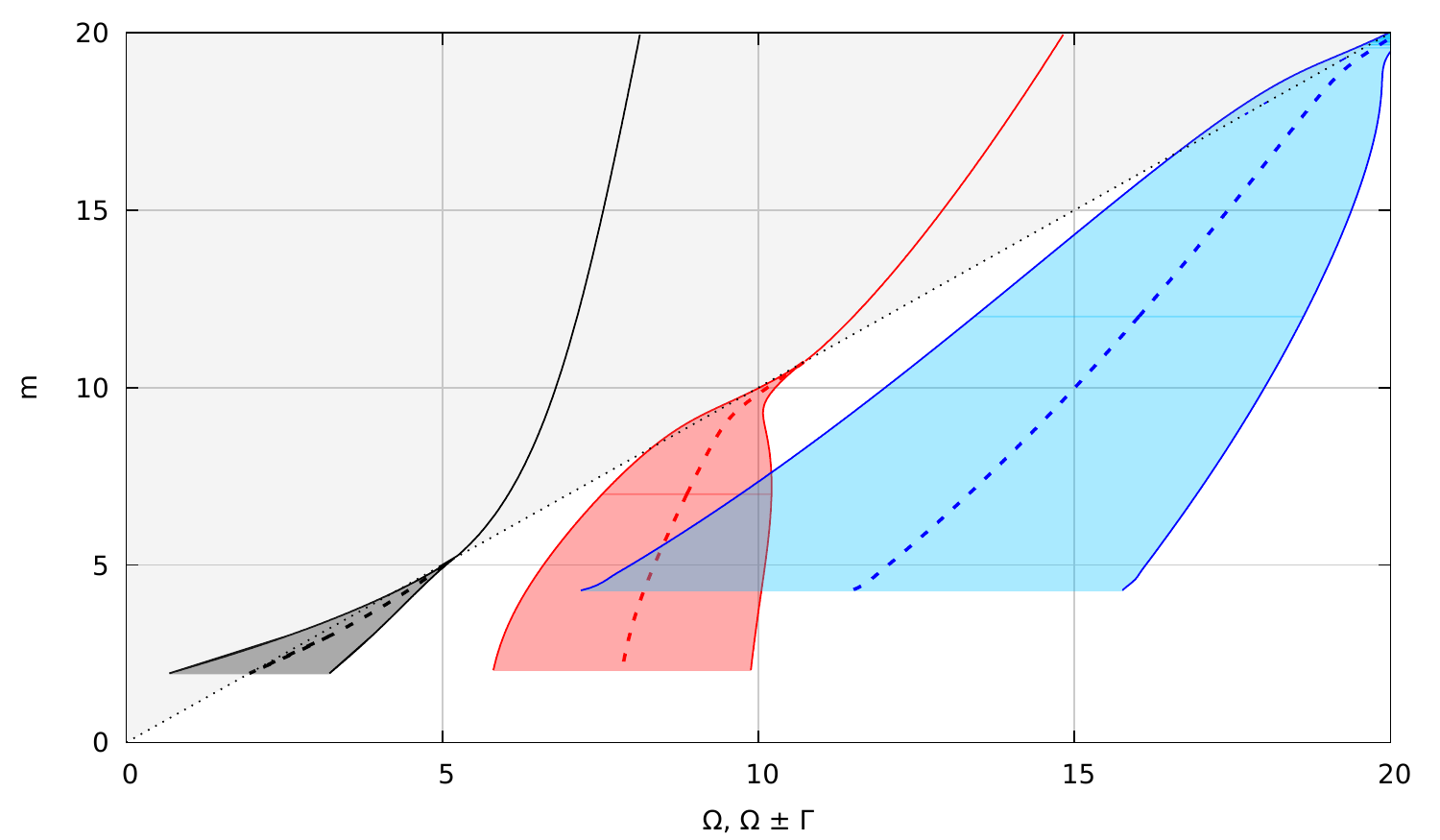}
\caption{Resonance modes for the unbinding potential with $\alpha=0.95$.}
\label{Res L024u}
\end{figure}
In the numerical computations we found three clear resonance modes when the value of the mass parameter equals the value used for small binding energies ($m=4.46$) 
\bea
\Omega_1+i\Gamma_1&=& 4.6077+0.1169 i \nonumber \\
\Omega_2+i\Gamma_2 &=&  8.2293+1.8342 i \nonumber\\
\Omega_3+i\Gamma_3 &=& 11.6761 + 4.1988 i 
\eea
The widths grows quite significantly. In particular, $\Gamma_1$ is much (more than ten times) smaller than $\Gamma_2$. So, for realistic parameter values the description of the Roper resonances is not satisfactory, again. Note, that for smaller value of the mass parameter the widths of the first two resonances go much closer and are,  in fact, almost identical for $m$ around 2 (see Fig. \ref{Res L024u}).
\subsection{Time dependence}
\begin{figure}
\includegraphics[width=0.9\textwidth,angle=0]{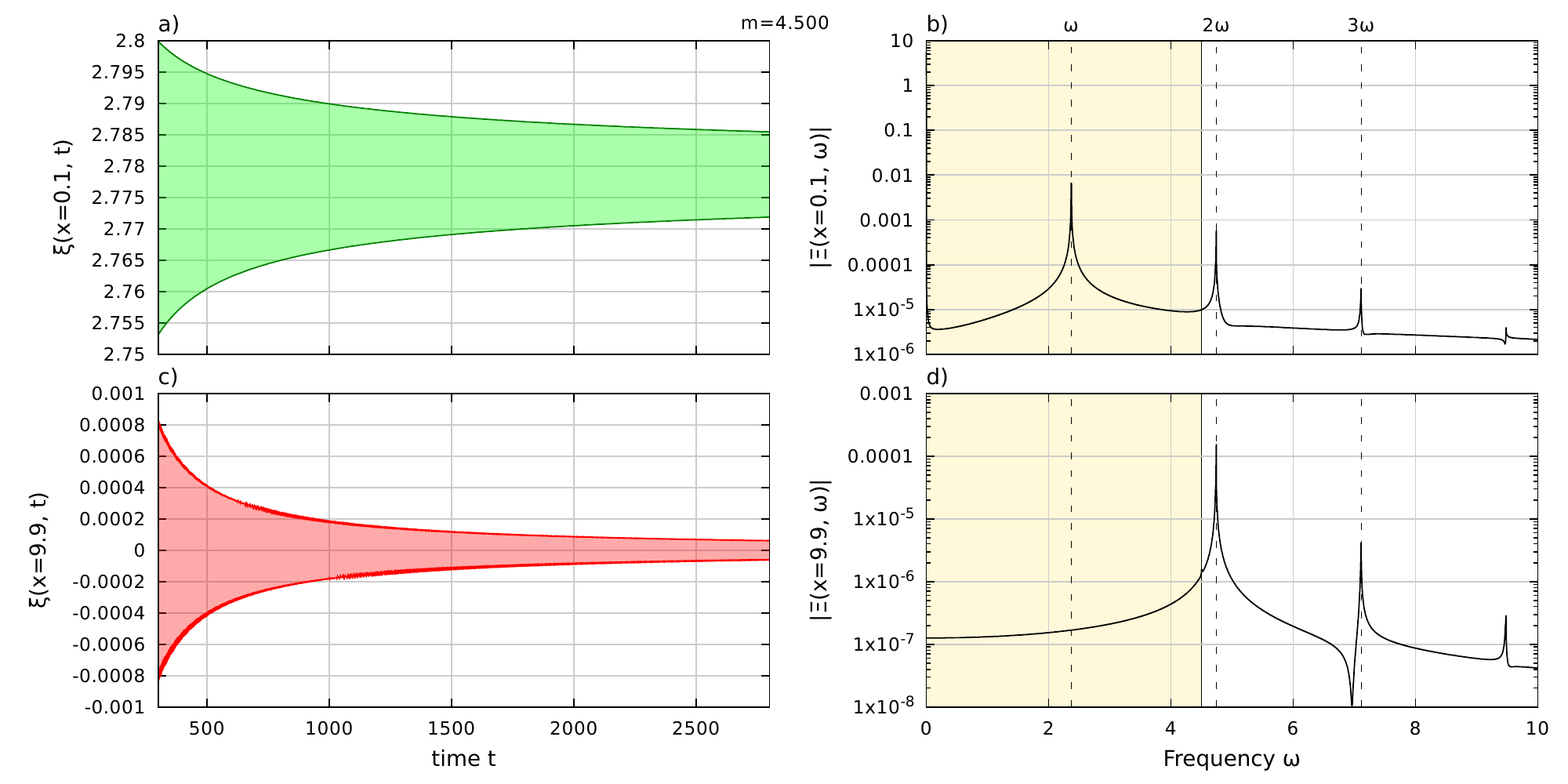}
\caption{\small Normal decay of the oscillational mode for $p=2$ case through the second harmonics.}\label{fig.normaldecay}
\includegraphics[width=0.9\textwidth,angle=0]{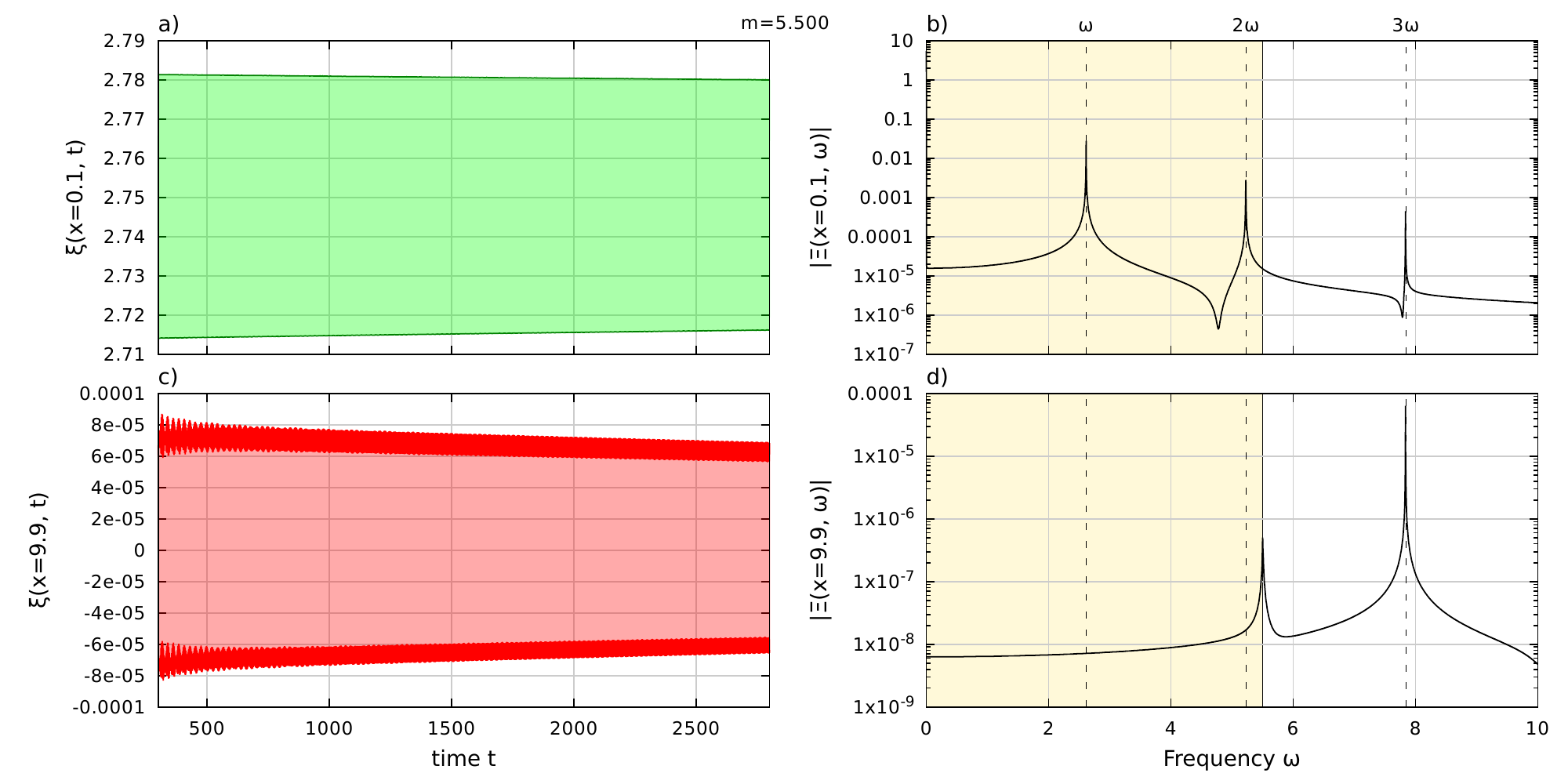}
\caption{\small Slow decay of the oscillational mode for $p=2$ case through the third harmonics.}\label{fig.slowdecay}
\end{figure}

In the linear regime, oscillational modes oscillate with constant amplitude and frequency. When nonlinear coupling is considered, the oscillational mode couples to higher harmonics. 
Usually already the second harmonics is above the mass threshold and propagates to spatial infinity, carrying away the energy from the oscillational mode. This results in the decay 
of the mode (see Fig. \ref{fig.normaldecay} for the $p=2$ Piette potential for $m=4.5$). However, for certain choices of parameters the second harmonics can still be below the mass threshold and only the third one propagates. 
Since higher harmonics are described by higher order perturbations, in this case the mode decays much slower than in the first case (see Fig. \ref{fig.slowdecay} for the $p=2$ Piette potential for $m=5.5$). This phenomenon was recently 
observed in \cite{DRS}. 

\section{The full Skyrme model and the role of the sextic term}
In this section, we consider the full Skyrme model where the sextic term is included. Moreover, we consider the family of potentials $\mathcal{U}^{(p)}$ introduced above. 
We remark that the resonance structure in the pure BPS Skyrme model and its dependence on a particular form of the potential was carefully investigated in \cite{BPS-res} (see also \cite{BPSvib2}).

From a qualitative point of view, the asymptotics is completely dictated by the potential contribution. In fact, $Q_\infty$ remains unchanged. Therefore, the existence of a mass threshold at which possible oscillating modes cease to exist continues to hold, as well. 

\begin{figure}
\includegraphics[height=6cm]{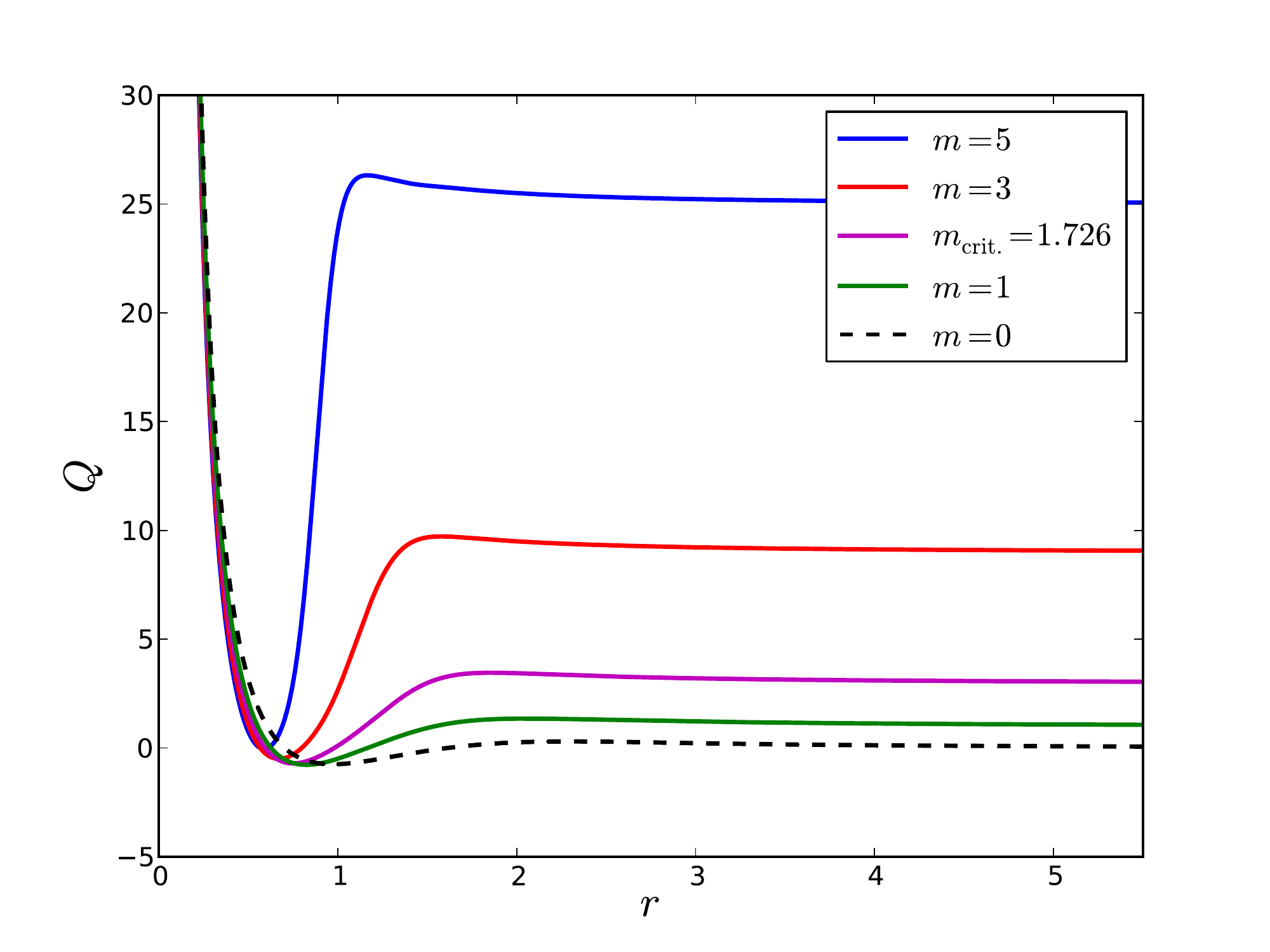}
\includegraphics[height=6cm]{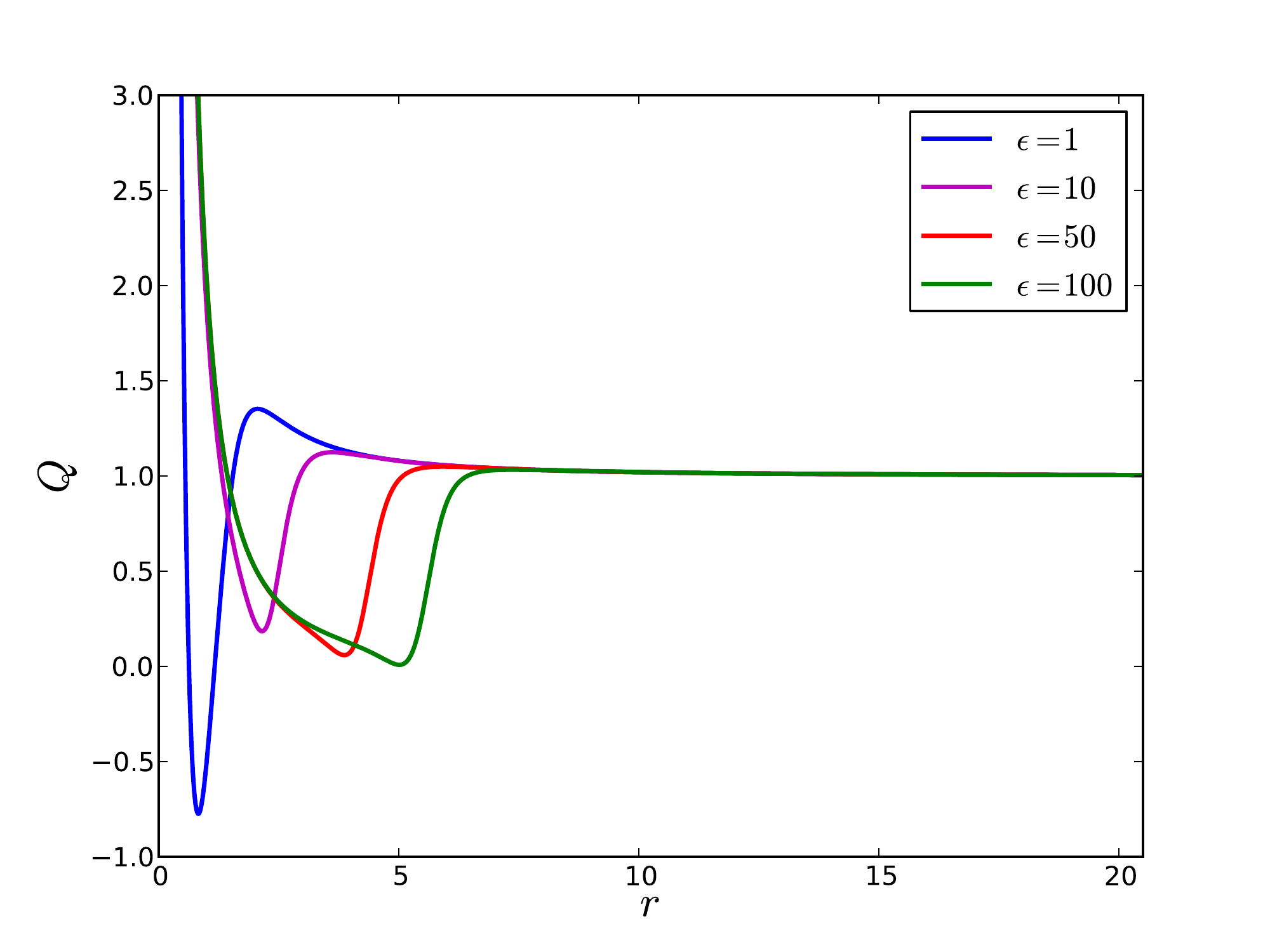}
\caption{Effective potential $Q$ for the $L_{2460}$ Skyrme model with the usual Skyrme potential. {\it Left:} $\epsilon=1$ and different value of the mass $m$. {\it Right:} $m=1$ and different values of $\epsilon$.} \label{Q full}
\end{figure}

\begin{figure}
\includegraphics[height=5cm]{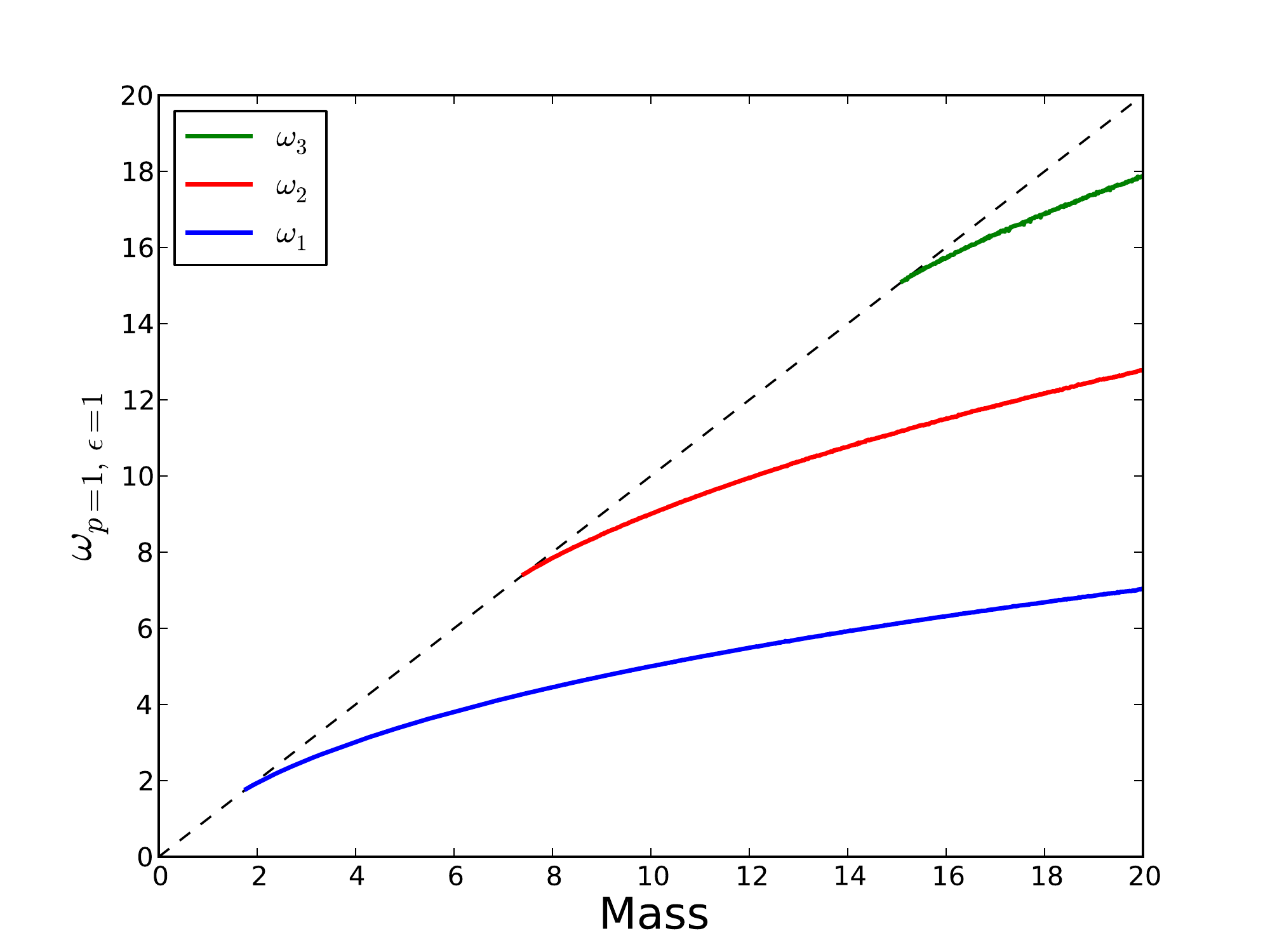}
\includegraphics[height=5cm]{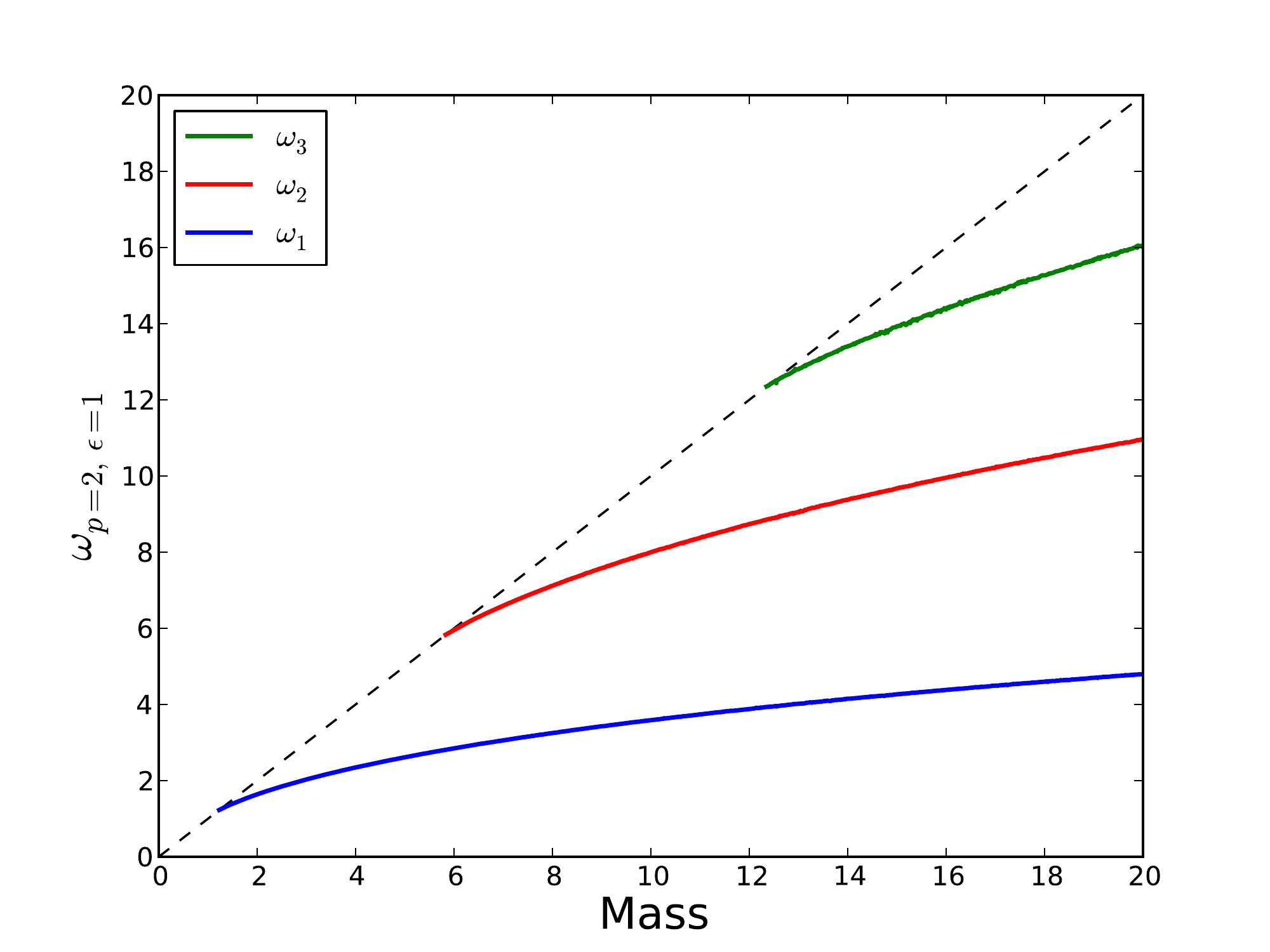}
\includegraphics[height=5cm]{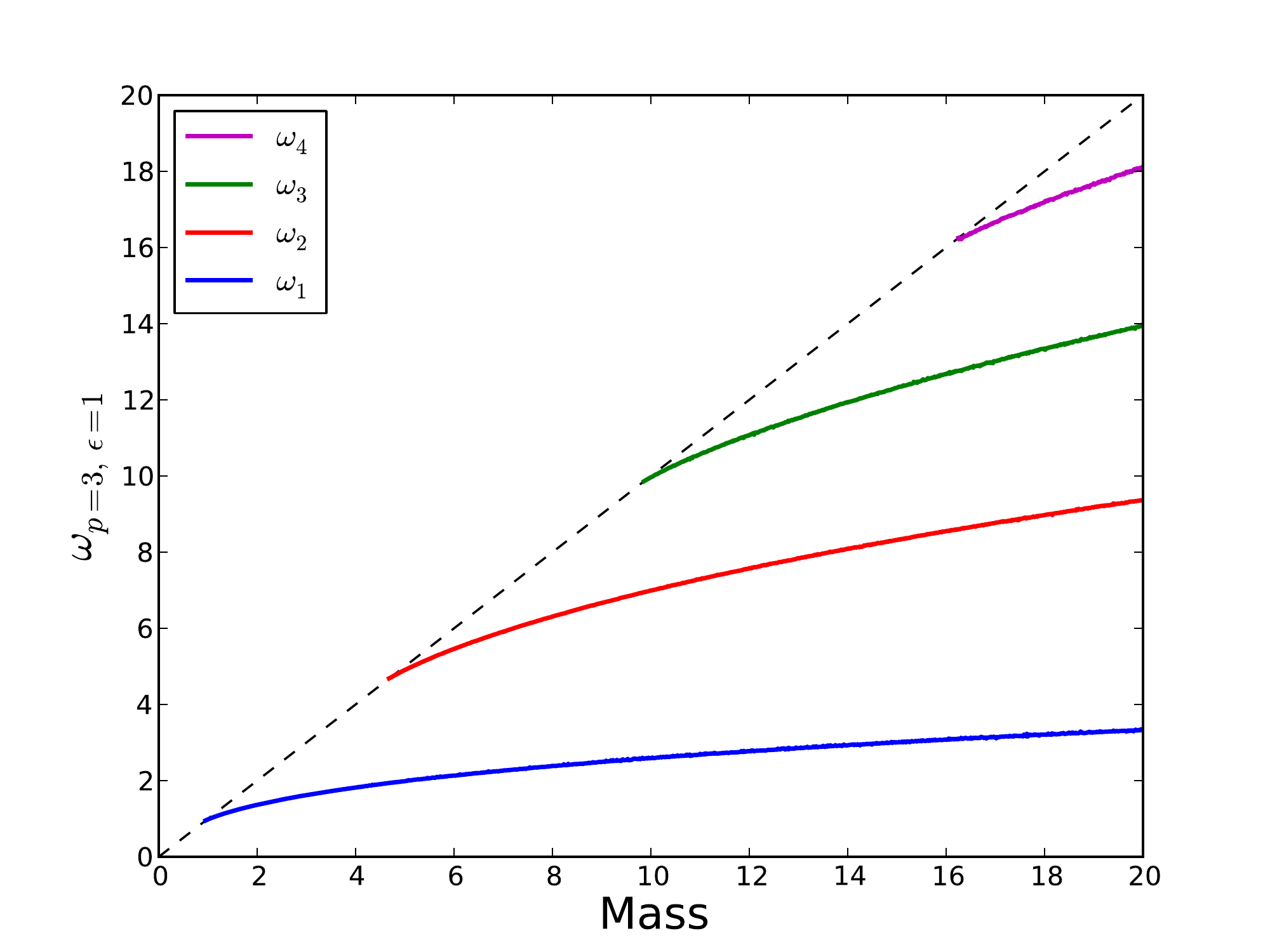}
\includegraphics[height=5cm]{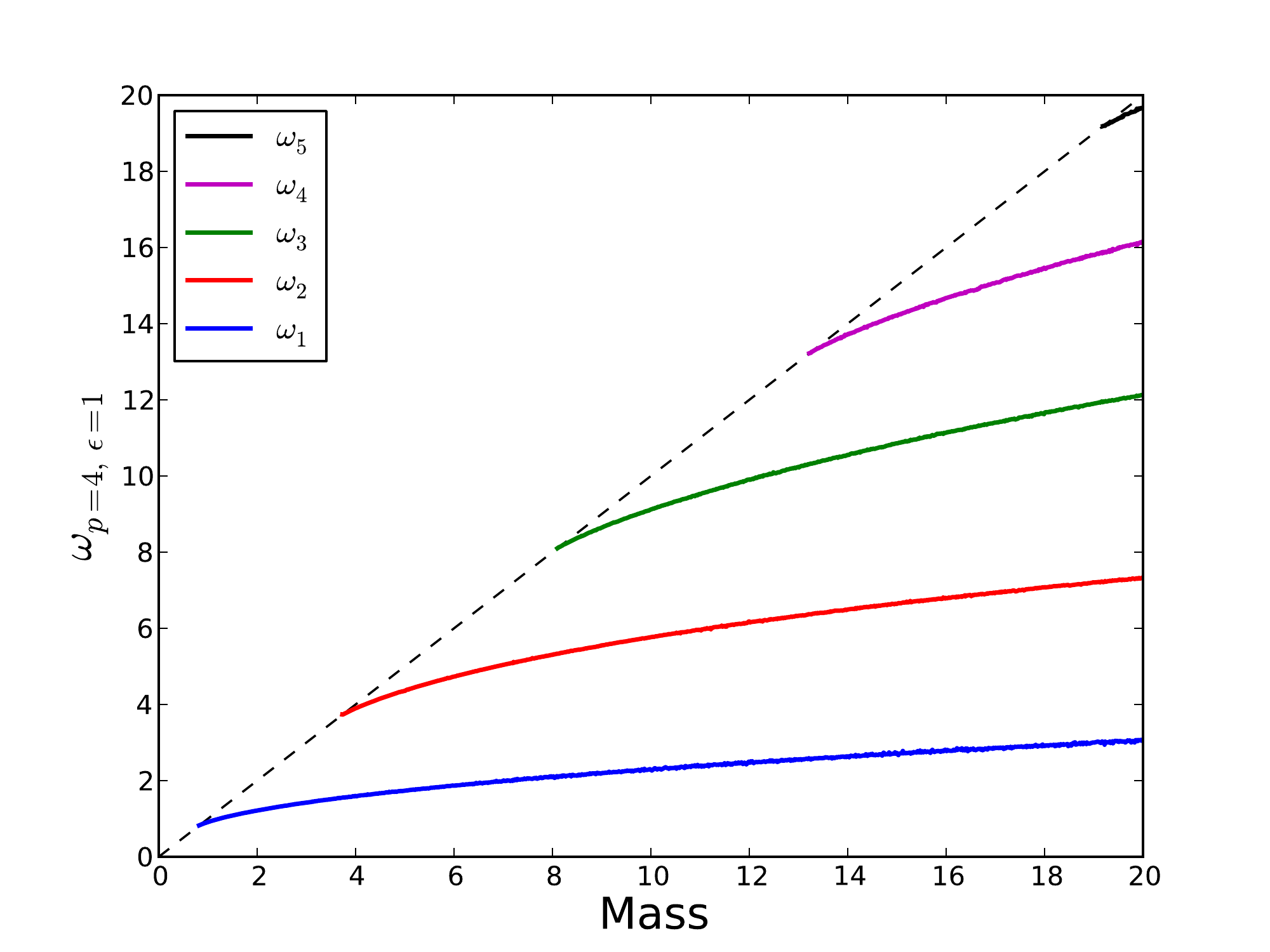}
\caption{The lowest oscillating modes for the $L_{2460}$ Skyrme model ($\epsilon=1$) with Piette's potential $\mathcal{U}^{(1)}$ as a function of the pion mass.} \label{omega_eps=1}
\end{figure}

\begin{figure}
\includegraphics[height=5cm]{E0246_omegas_mass0to20_p1_eps1.pdf}
\includegraphics[height=5cm]{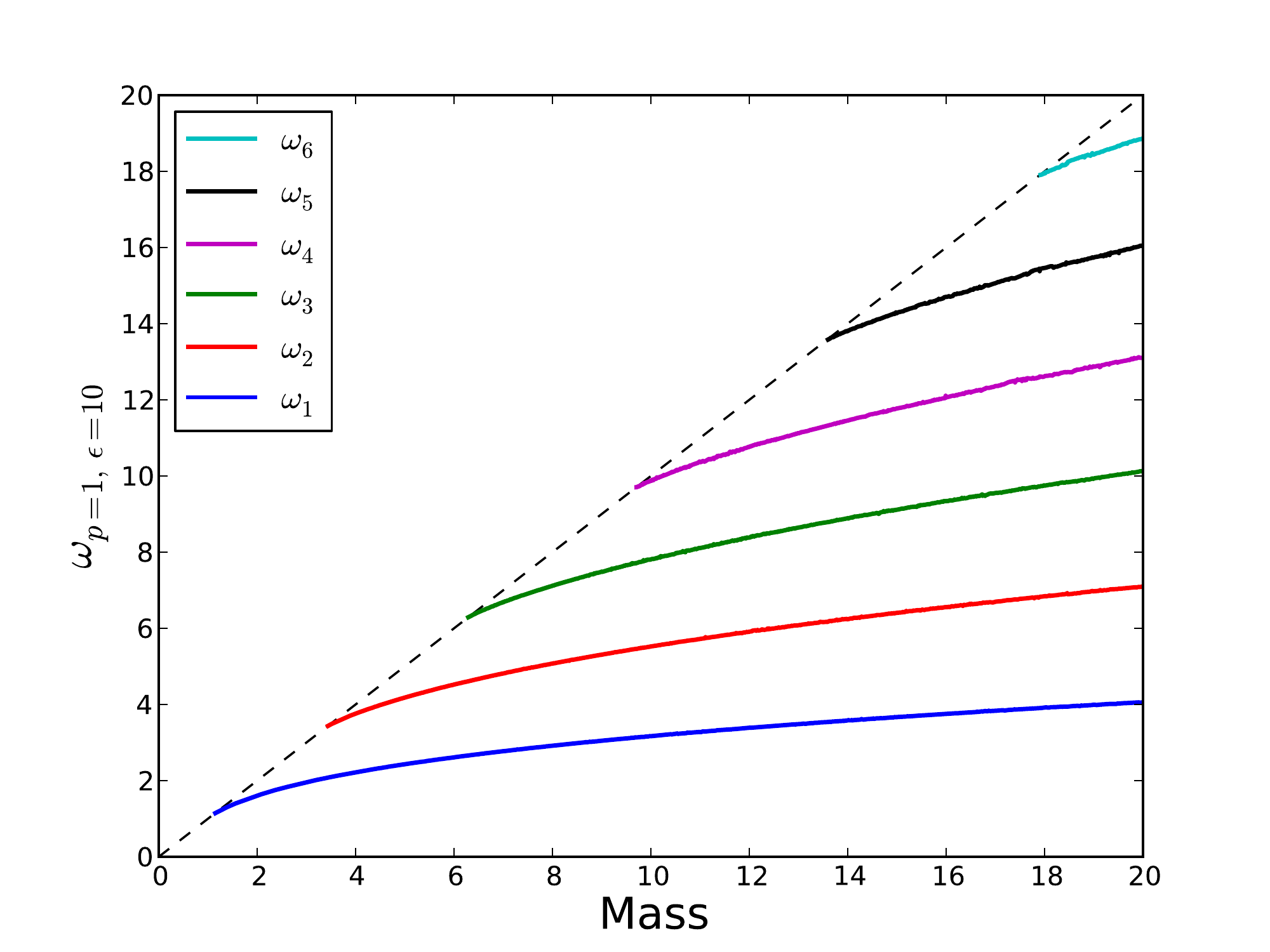}
\includegraphics[height=5cm]{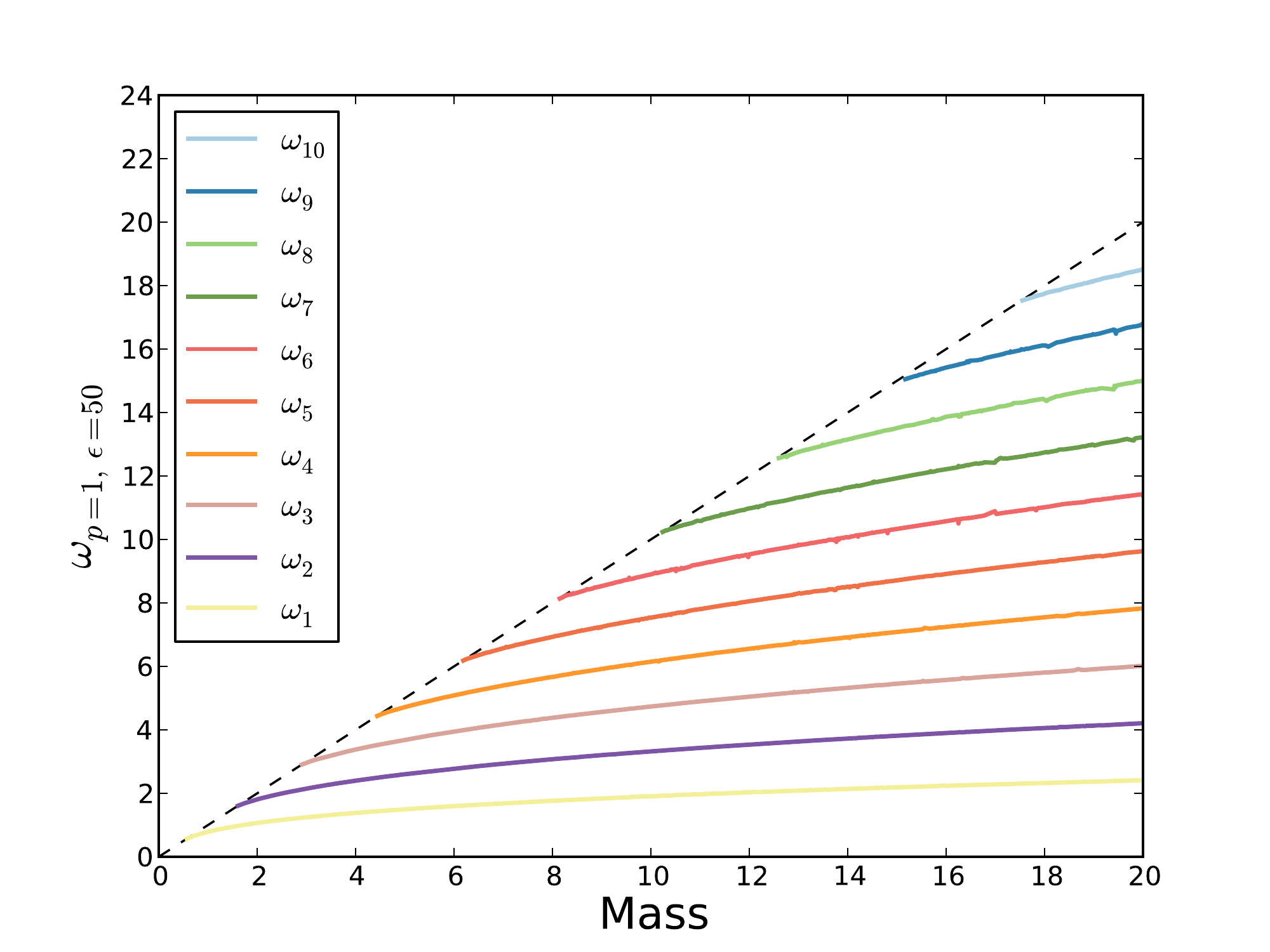}
\includegraphics[height=5cm]{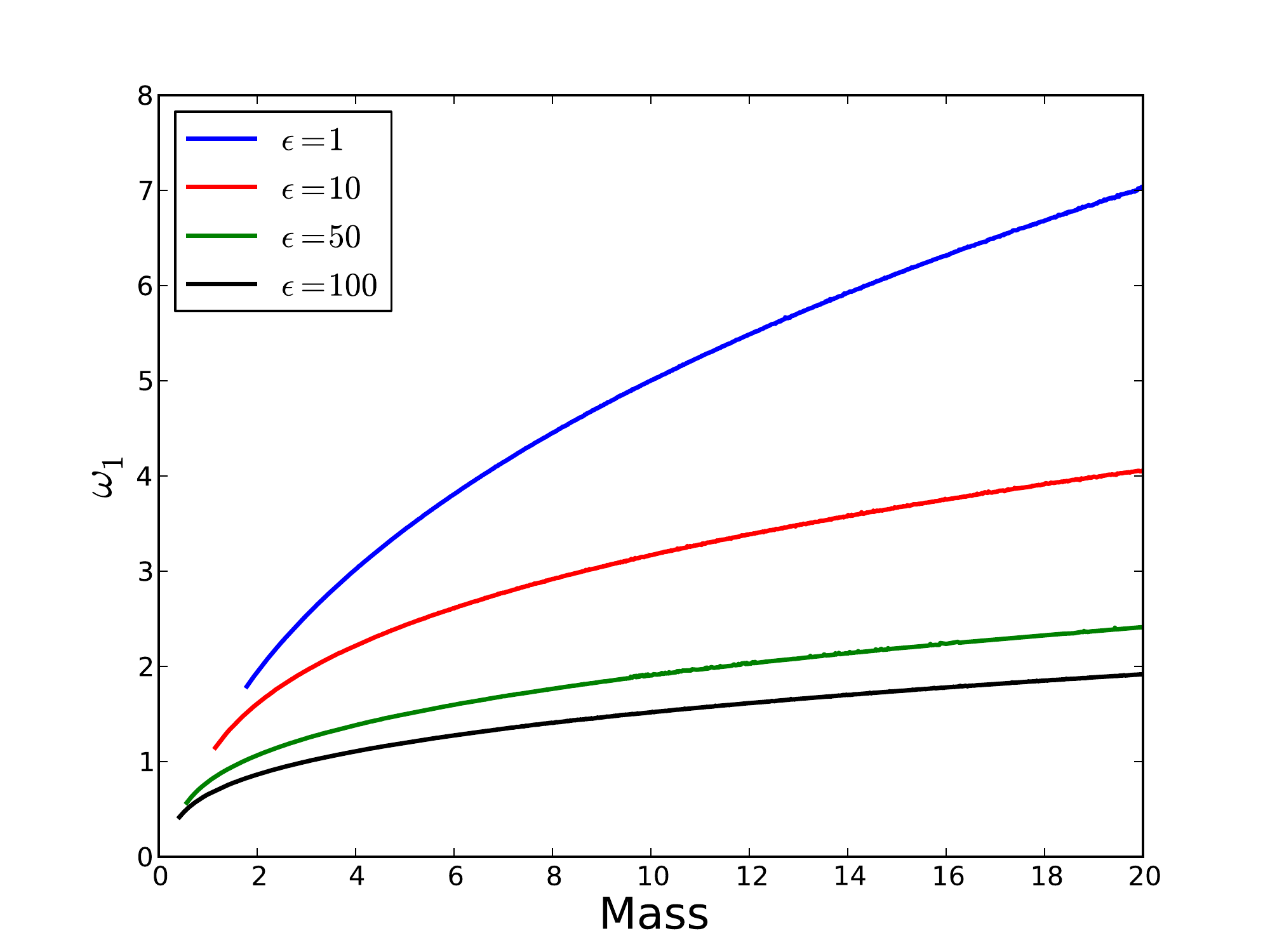}
\caption{The lowest oscillating modes for the $L_{2460}$ Skyrme model with the usual Skyrme potential $\mathcal{U}^{(1)}_{p=1}$ for $\epsilon=1,10,50$.} \label{omega_eps_1_50}
\end{figure}

\begin{figure}
\includegraphics[height=5.cm]{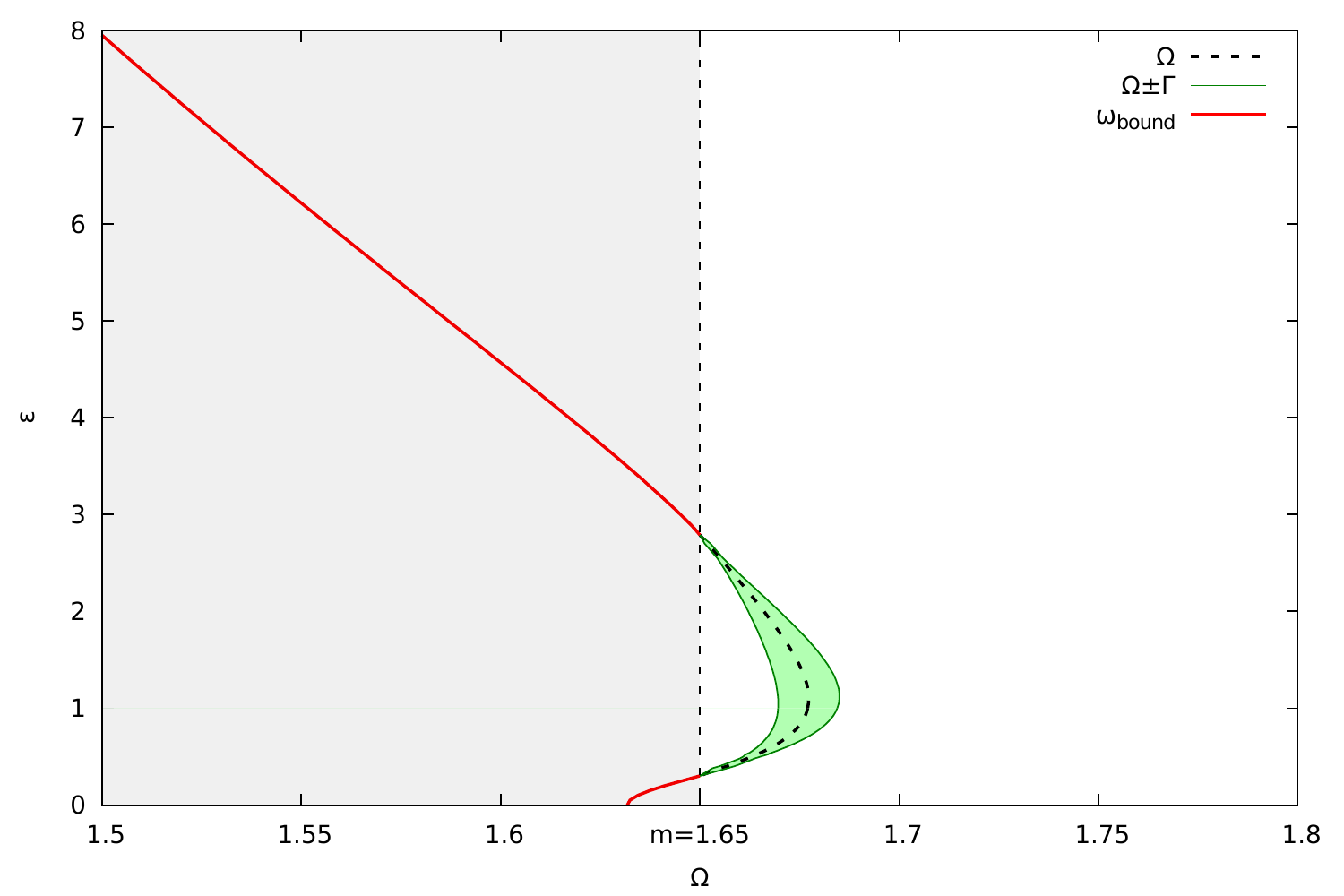}
\includegraphics[height=5.cm]{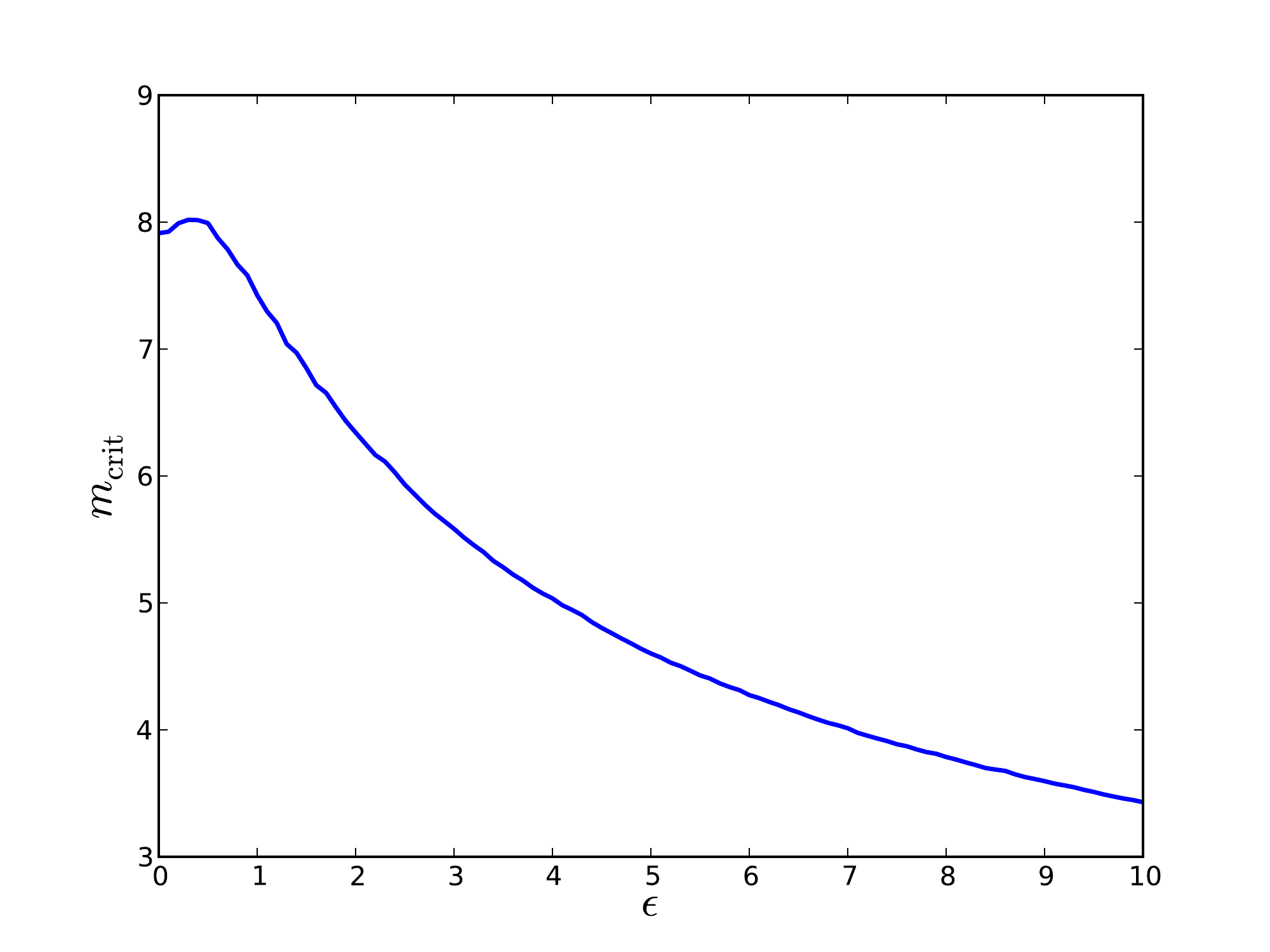}
\caption{{\it Left:} The first oscillating/resonance mode for $m=1.65$. {\it Right:} Value of the critical mass for the second oscillating mode} \label{m crit}
\end{figure}

\begin{figure}
\hspace*{-1.0cm}
\includegraphics[height=9.5cm]{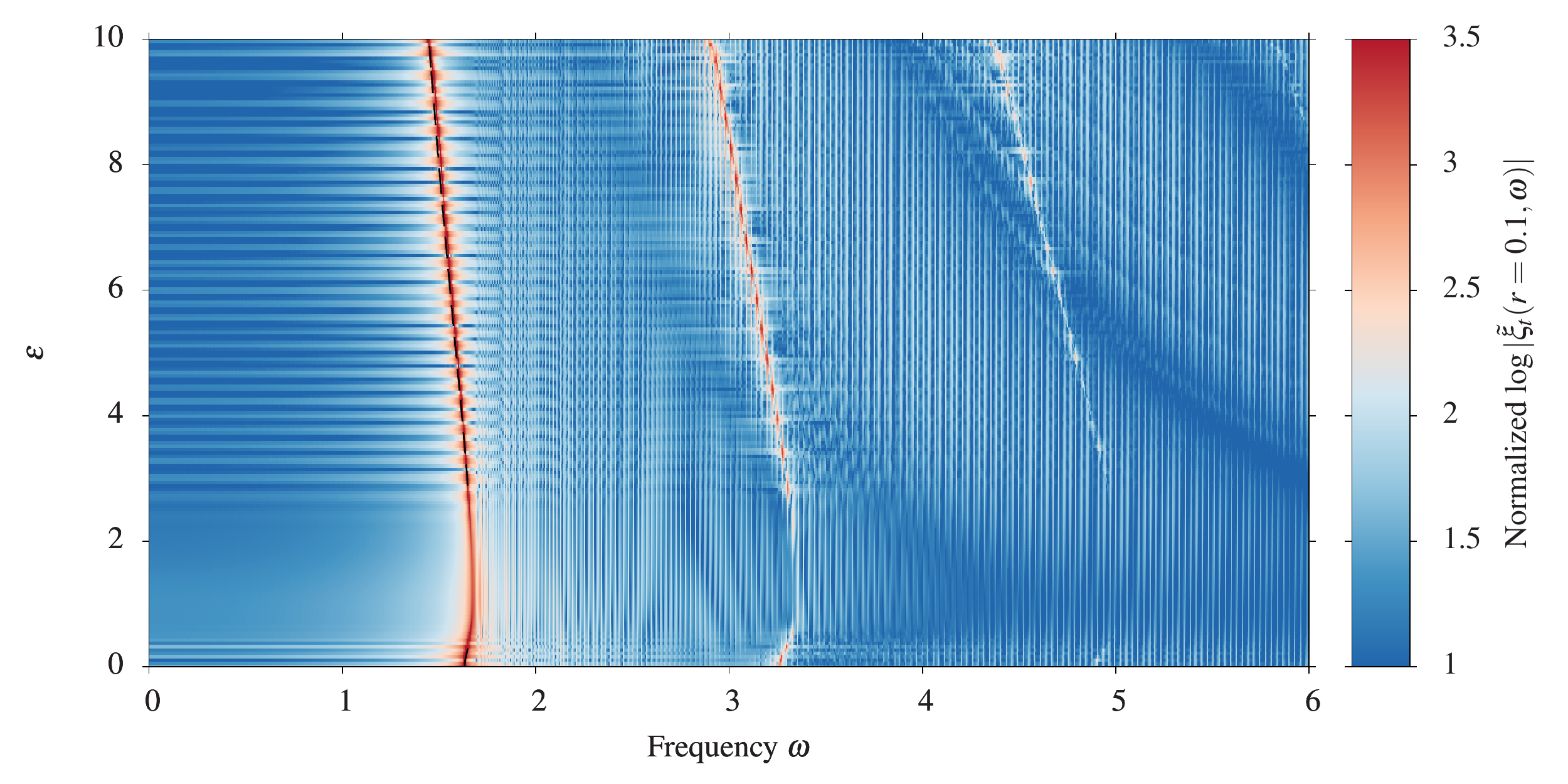}
\caption{FFT scan of the full dynamics for the full Skyrme model $\mathcal{L}_{0246}$ with the usual potential and $m=1.65$. A window 
where the first oscillating mode is replaced by a resonance is visible. The dashed line is the oscillating mode.}
\label{scan_L_0246}
\end{figure}

\begin{figure}
\hspace*{-1.0cm}
\includegraphics[height=5.0cm]{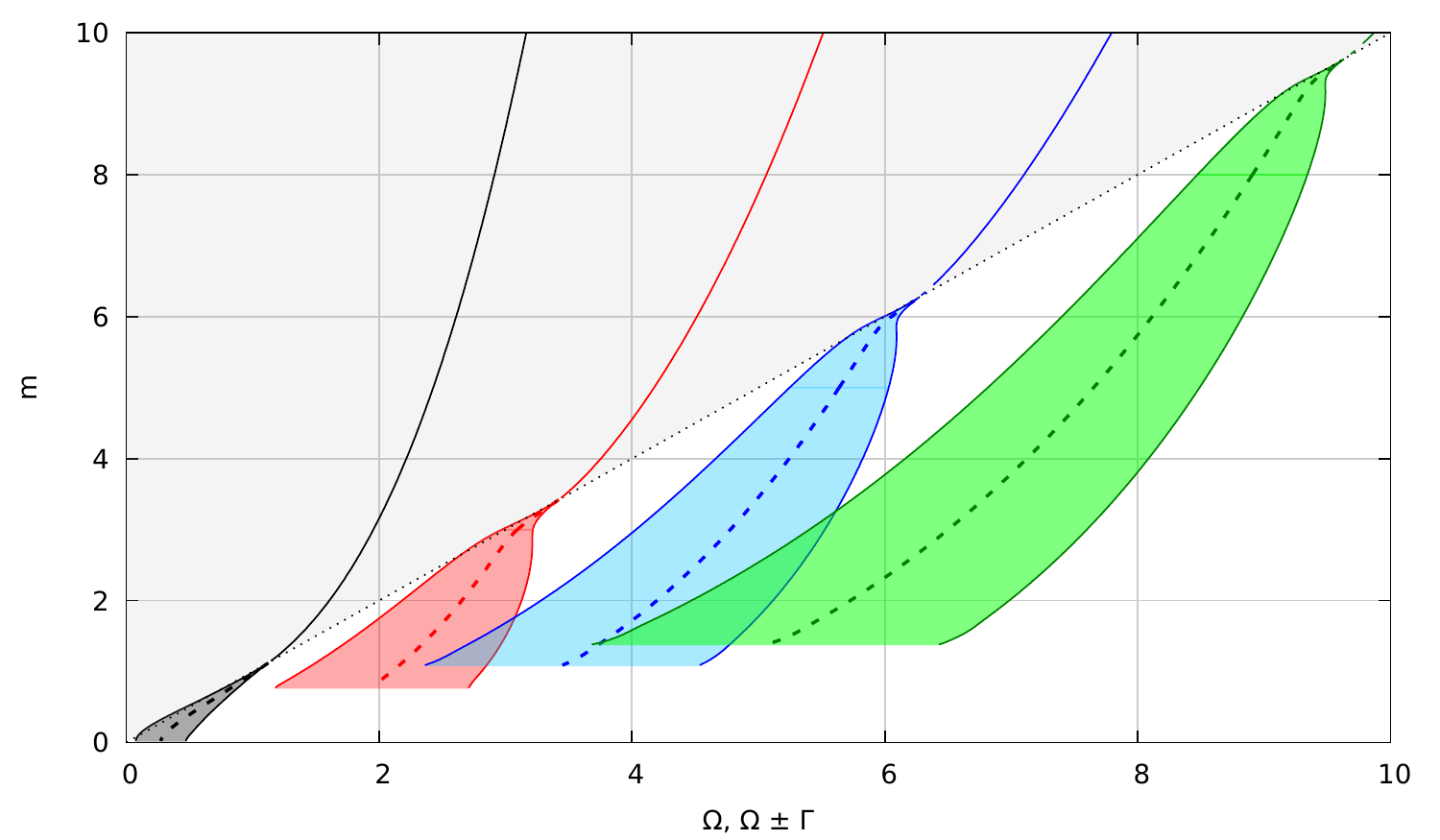}
\caption{Resonance modes for the full Skyrme model with $\epsilon=10$ and the usual pionic potential.}
\label{Res L0246}
\end{figure}

In Fig. \ref{Q full} (right panel) we show the effective potential in the case of the usual Skyrme model and for several values of the $\epsilon$ constant (while other couplings are fixed). Qualitatively, the inclusion of the sextic term to the usual Skyrme model (with the previously chosen values of the parameters) leads to two effects: 1) the bottom of the little well is raised above 0. This happens at least for sufficiently large $\epsilon$; 2) the width of the well is widened. The first effect means that the sextic term prevents the appearance of a negative energy bound state and therefore contributes to the stability of the Skyrmion. The second effect results in lowering the value of the critical mass below which there is no oscillating mode, and in increasing the number of oscillating modes - Fig. \ref{omega_eps=1}. Furthermore, the  resonance modes become narrower. 
These two effects, which are visible for higher values of $\epsilon$ (see for example Fig. \ref{omega_eps_1_50}) do not occur for small perturbations of the $\mathcal{L}_{024}$ model. Indeed, for small $\epsilon$ (below 1) the critical mass for the first oscillating mode grows, which means that inclusion of a small fraction of the sextic term can lead to the disappearance of some oscillating modes (and the appearance of resonance modes) - see Fig. \ref{m crit}. The equivalent result for the full numerics is seen in Fig. \ref{scan_L_0246}. The linear resonance spectra for different pion mass values are shown in Fig. \ref{Res L0246}.
As we see, the inclusion of the sextic term has a manifold and quite nonlinear impact on the structure of the quasi-normal modes. Furthermore, for the chosen potentials, the differences between the widths of the first few resonances, again, are too big to reproduce the experimental data.


\begin{table}
\begin{center}
\begin{tabular}{c|ccccc}
\hline
\hline
 & $n=1$ & $n=2$ &  $n=3$&  $n=4$ & $n=5$ \\
\hline 
\hline
$\epsilon=0 \;\;\;$  &  1.59 & 7.98 &19.37  & - &- \\
\hline
$\epsilon=1 \;\;\;$ &  1.79 & 7.42 & 15.12 & -&-  \\
\hline
$\epsilon=10 \;\;\;$ &  1.14 & 3.43 & 6.29 & 9.69 & 13.52  \\
\hline
$\epsilon=50 \;\;\;$ &  0.56 & 1.60 & 2.90 & 4.44&6.18  \\
\hline
\hline
\end{tabular}  
\caption{Critical masses for the first five oscillating modes for the full Skyrme model with the usual Skyrme potential. $\mathcal{U}_{p=1}^{(1)}$. No value denotes that the critical point occurs above $m=20$.} \label{table L0246}
\end{center}
\end{table}
\begin{table}
\begin{center}
\begin{tabular}{c|ccccc}
\hline
\hline
 & $n=1$ & $n=2$ &  $n=3$&  $n=4$ & $n=5$ \\
\hline 
\hline
$p=1 \;\;\;$  &  1.59 & 7.99 &19.38  & - &- \\
$p=2 \;\;\;$ &  1.18 & 6.47 & 16.22 & -&-  \\
$p=3 \;\;\;$ &  0.95 & 5.07 & 12.38 & -&-  \\
$p=4 \;\;\;$ &  0.85 & 3.98 & 9.77 & 17.56&-  \\
\hline
\hline
\end{tabular}  
\hspace*{0.2cm}
\begin{tabular}{c|ccccc}
\hline
\hline
 & $n=1$ & $n=2$ &  $n=3$&  $n=4$ & $n=5$ \\
\hline 
\hline
$p=1 \;\;\;$  & 1.79 & 7.42 & 15.12 & -&-  \\
$p=2 \;\;\;$ &  1.23 & 5.84 & 12.36 & -&-  \\
$p=3 \;\;\;$ &  0.96 & 4.69 & 9.86 & 16.25&-  \\
$p=4 \;\;\;$ &  0.83 & 3.74 & 8.10 & 13.23&19.19  \\
\hline
\hline
\end{tabular}  
\caption{Critical masses for the full Skyrme model with the Piette potentials. No value denotes that the critical point occurs above $m=20$. {\it Left:} $\epsilon=0$. {\it Right:} $\epsilon=1$.} 
\end{center}
\end{table}

\newpage
\section{Reconstruction}
Up to now we have been considering models with a given potential $\mathcal{U}(\xi)$. We found static Skyrmionic solutions in the charge one sector. Then we perturbed the static solutions and obtained the effective potential $Q(r)$ which is the core ingredient in the linear perturbation theory. Finally, we analysed the spectral structure of the potential $Q(r)$ by solving the appropriate Sturm-Liouville equation, obtaining resonances or bound states. Obviously, qualitative as well as quantitative properties of the spectra are determined by a particular form of the effective potential $Q(r)$. 

Since there are many acceptable potentials in the Skyrme framework, provided they lead to the physical pion mass and reasonable binding energies, one can pose the opposite problem. Namely, for a given (qualitatively or even quantitatively reasonable) vibrational spectrum, which results from a given effective potential $Q(r)$, to reconstruct the original potential $\mathcal{U}$. In this section we show how from the linearised potential $Q(r)$ the full nonlinear model can be obtained.

\subsection{$1+1$ dimensional example}
First we start with a simple example of one scalar field in $1+1$ dimensions.  Then, the equation for the static solution is 
\begin{equation} \label{V1}
 -\xi_0''+\mathcal{U}_\xi(\xi_0)=0.
\end{equation} 
The linearised equation describing small perturbations around the static solution can be found as
\begin{equation}
 -\eta''+Q(x)\eta=\omega^2\eta,
\end{equation} 
where 
\begin{equation} \label{V3}
 Q(x) = \mathcal{U}_{\xi\xi}(\xi_0(x)).
\end{equation} 
For concreteness we assume 
\begin{equation}
 Q(x)=m^2-\frac{2}{\cosh^2x}.
\end{equation} 
Introducing a new variable $\mathcal{W}=\mathcal{U}_\xi$, we can write the following system of equations for $\mathcal{W}$ and $\xi$, equivalent to Eqs. (\ref{V1}) and (\ref{V3}):
\begin{equation}
 \begin{cases}
  \mathcal{W}_\xi=Q,\\
  \xi_{xx}=\mathcal{W}.
 \end{cases}
\end{equation} 
After changing the variable  $\mathcal{W}_\xi=\mathcal{W}_x/\xi_x$ the system can be written as
\begin{equation}\label{eq:system1}
 \begin{cases}
  \mathcal{W}_x=Q\xi_x,\\
  \xi_{xx}=\mathcal{W}.
 \end{cases}
\end{equation} 
Formally, the system can be solved after applying appropriate boundary conditions. Namely, $\xi$ has to satisfy the usual topological boundary conditions. For 
symmetric potentials we can assume that $\xi(0)=\pi$, which is the maximum of the potential (topological zero) and $\xi(\infty)=\xi_{vac}=0$ which is 
the real vacuum (minimum of the field theory potential). Since both of these points are extrema of the potential, $\mathcal{U}_\xi=\mathcal{W}=0$ at these points. The problem to solve is a two point boundary 
condition problem. 
Starting from $x=0$ we have to satisfy two conditions at $x=\infty$. To do that, two (shooting) parameters have to be used. One of them is $\xi_x(0)$. 
The second is the choice of $m$.\\
In this trivial example, the second equation can be differentiated and we obtain
\begin{equation}
 \xi_{xxx}=\mathcal{W}_x=Q\xi_x
\end{equation} 
This is the linearised equation for $\eta=\xi_x$ with $\omega=0$. Actually, $\xi_x$ has an interpretation as the translational mode of the soliton 
$\xi(x+a)\approx\xi(x)+a\xi_x(x)+\mathcal{O}(a^2)$. $\xi_x$ is a small perturbation, and since the model has the translation symmetry, $\xi+a\xi_x$ is 
also a static solution, so $\omega$ must be zero. In our working example, the potential has a single bound mode $\eta(x)=1/\cosh(x)$ with the frequency 
$\omega^2=m^2-1$. So $m=1$.
Now we have to solve the equation $\xi_x=c_1/\cosh(x)$, which fortunately has a closed form
\begin{equation}
 \xi(x)=2c_1\tan^{-1}\left(\exp(x)\right)+c_2.
\end{equation} 
Applying boundary conditions we obtain $c_1=-2$, $c_2=2\pi$. Now we know the profile of the static soliton solution. From the equation 
$\mathcal{W}=\xi_{xx}=\mathcal{U}_\xi$ we can obtain:
\begin{equation}
2\xi_x\xi_{xx}=2\xi_x\mathcal{U}_\xi \quad\Rightarrow\quad\mathcal{U}_x=\frac{1}{2}\left(\xi_x^2\right)_x\quad\Rightarrow\quad 
\mathcal{U}(x)=\frac{1}{2}\xi_x^2+C.
\end{equation} 
The last equation can be treated as an implicit (parametric) equation for the potential $\mathcal{U}(\xi)$. The final form can be untangled by using the relation that 
$\frac{1}{2}\xi_x^2=1+\cos\xi$ and setting $C=0$. The obtained example is in fact the shifted sine-Gordon model
\begin{equation}
 \mathcal{U}(\xi) = 1-\cos\xi.
\end{equation} 
Similarly, from the potential 
\begin{equation}
 Q=m^2-\frac{6}{\cosh^2(x)}
\end{equation} 
the well known $\phi^4$ model can be obtained. 

The working example was a simple $1+1d$ model which can be easily separated. However, we have also tested the procedure by solving numerically 
the system (\ref{eq:system1}) with the boundary conditions $\mathcal{W}(0)=\mathcal{W}(\infty)=\xi(\infty) = 0$ and $\xi(0)=\pi$.

\subsection{Skyrme model from the linearised potential}
\begin{figure}
\includegraphics[width=0.9\textwidth,angle=0]{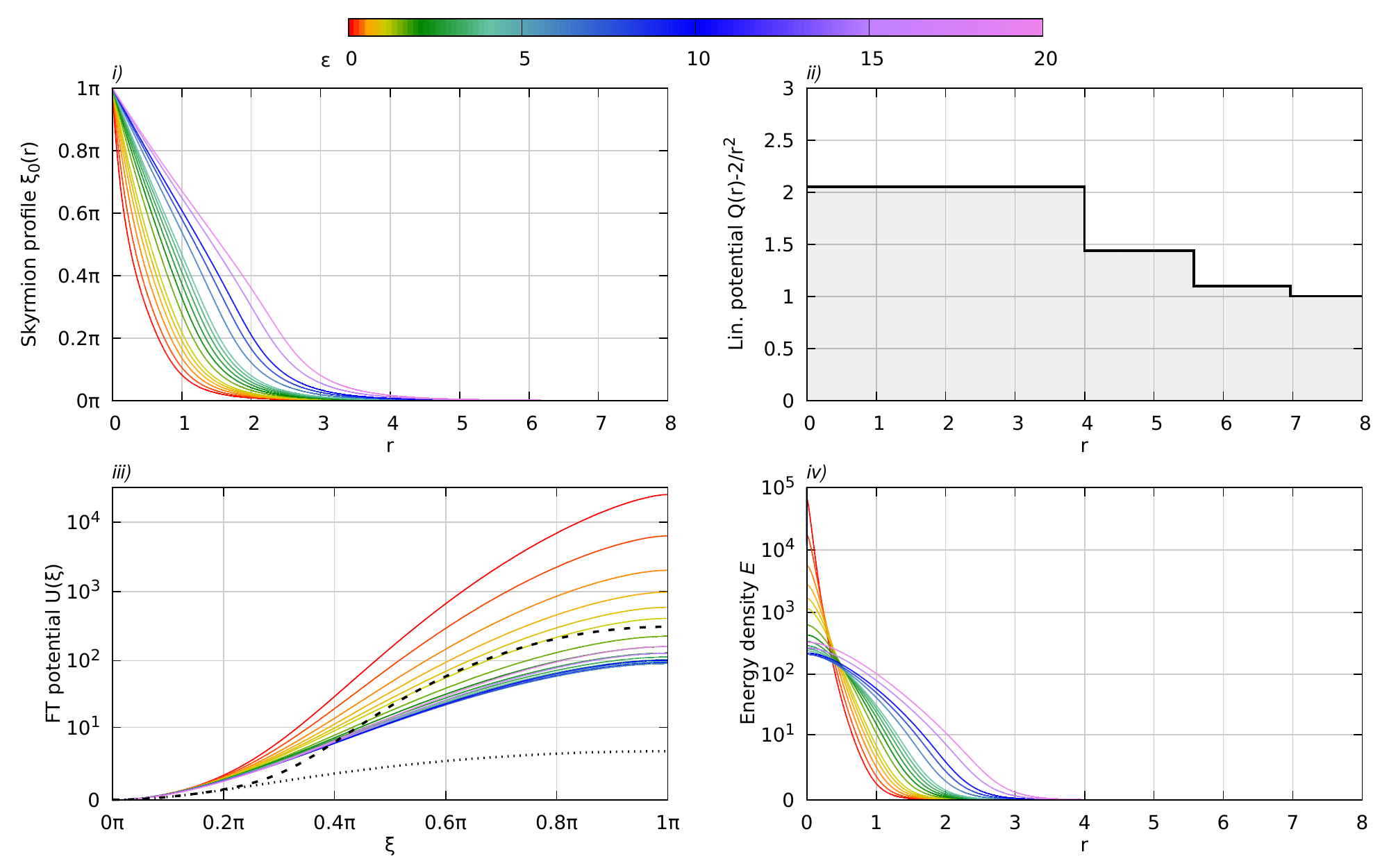}
\caption{\small Reconstruction of the field theoretic potential $\mathcal{U}$ for the full Skyrme model $\mathcal{L}_{0246}$. {\it (i)} profile of the solution; {\it (ii)} field theoretical potential $\mathcal{U}$; {\it (iii)} the effective potential $\mathcal{Q}$ with the 3-dimensional universal repulsive core subtracted; {\it (iv)} the energy density. The structure of resonances is described in the text.}\label{rec}
\end{figure}

In principle, we could repeat the procedure described in the previous subsection for Skyrmions and obtain $\mathcal{U}$ from a linearised potential $Q$ with the desired spectral properties. However, the procedure requires the solution for the translational mode. Our original theory is translation invariant, but our hedgehog ansatz is not. Moreover, we don't expect that at any point we could find a closed form of the potential, therefore, we focus on the numerical procedure rather than on the analytical approach. 

The equations to be solved are the following 
\begin{enumerate}
 \item The equation for the static solution $\xi_0$ of Eq. (\ref{stat-ode})
 \item  $ Q=Q_0(r,\xi)+Q_\mathcal{U}(r, \xi, \mathcal{W}, \mathcal{W}_{\xi})$ where $\mathcal{W} = \mathcal{U}_\xi$.
\end{enumerate}

In Fig. \ref{rec} we present an example of the (numerical) reconstruction procedure for the effective potential assumed in the form of a potential well (with the usual dimensionally induced $\frac{2}{r^2}$ core). By numerical integration of the problem defined above we found the corresponding potential $\mathcal{U}$. 

Now, the strategy is obvious. We construct $Q$ which leads to narrow resonance modes with comparable widths and repeat the reconstruction. For example we assume the following structure of resonances:
\bea
\Omega_1+i\Gamma_1&=& 1.710 + 0.250 i \nonumber \\
\Omega_2+i\Gamma_2 &=&  1.880+  0.250 i \nonumber\\
\Omega_3+i\Gamma_3 &=& 2.384  +0.228 i \nonumber \\
\Omega_4+i\Gamma_4 &=& 2.869 +  0.336 i 
\eea
where the main choice is dictated by similar widths of the resonances. The corresponding field theoretical potential is very peaked at the anti-vacuum which corresponds to a significant (probably unphysical) concentration of the energy density at the origin. This unpleasant fact is cured if we increase the value of the coupling constant multiplying the sextic term, i.e., once we move the model towards the near-BPS Skyrme model. Then, the energy density becomes flatter. 

\subsection{Roper reconstruction details}
Here, we use the reconstruction procedure described in the previous section for parameter families of linear potentials $Q$, and fit these parameters to physical observables of the nucleon, like the Roper resonances, the nucleon mass and the isoscalar electric charge radius of the nucleon.
For the reconstruction we have decided to use piecewise flat potentials (plus the usual $2/r^2$ term)
because the resulting problem can be solved partly analytically. The potential can then be written in the form 
\begin{equation}
 Q(r; \{Q_n,L_n\}) = \frac{2}{r^2} + \sum_{n=0}^{N-1} Q_n H(r-r_n)H(r_{n+1}-r)
\end{equation} 
where $\{Q_n,L_n\}$, $n=0, \ldots N-1$, are the parameters of the potential, the height and length of each segment,
\begin{equation}
 r_n=\sum_{i< n}L_i,
\end{equation} 
and $H(x)$ is the Heaviside step function. Further, $r_0 = 0 < r_1 < r_2 < {\ldots} < r_N = \infty$, the potential consists of $N$ steps, and there are $N-1$ matching points $r_1 <r_2 <\ldots r_{N-1}$. \\
The solutions are known
\begin{equation}\label{eq:decomp2}
 u_n(r)=A_nu_{1,n}(r)+B_nu_{2,n}(r)\, , \qquad\text{where}\qquad r_n \leq r \leq r_{n+1}
\end{equation} 
and
\begin{equation}
 u_{1,n}(r) = \left(1+\frac{i}{k_nr}\right)e^{ik_nr}\qquad\text{and}\qquad u_{2,n}(r)=\left(1-\frac{i}{k_nr}\right)e^{-ik_nr},\qquad k_n=\sqrt{\omega^2-Q_n}.
\end{equation} 
The solutions should have the same values and derivatives at the matching points. The matching can be performed with the help of the transition matrix defined as ($n=1,\ldots ,N-1$)
\begin{equation}
 M(k_{n}, k_{n-1}, r_n)=
\begin{bmatrix}
  u_{1,n}& u_{2,n}\\
  u'_{1,n}& u'_{2,n}
\end{bmatrix}^{-1}_{r_n}
\begin{bmatrix}
  u_{1,n-1}& u_{2,n-1}\\
  u'_{1,n-1}& u'_{2,n-1}
 \end{bmatrix}_{r_n} ,
\end{equation} 
which transfers the solution from one segment to the next,
\begin{equation}
 \begin{bmatrix}
  A_n\\B_n
 \end{bmatrix}=
  M(k_{n}, k_{n-1}, r_n)
  \begin{bmatrix}
  A_{n-1}\\B_{n-1}
 \end{bmatrix} .
\end{equation} 
The asymptotic solution can be found by simply multiplying these matrices in the appropriate order (larger $n$ on the lhs) and acting on the initial conditions
\begin{equation}
 \begin{bmatrix}
  A_N\\B_N
 \end{bmatrix}=
 \prod_{n=N}^1 M(k_{n}, k_{n-1}, r_n)
 \begin{bmatrix}
  A_0\\B_0
 \end{bmatrix}
\end{equation}

The resonance condition $A_N/B_N=0$ for $A_0=B_0=1$ is just an algebraic equation for $\omega$. 
Solving such an equation numerically is much simpler than finding the resonance modes integrating the appropriate ODEs.
Using the procedure described in the previous section, we can reconstruct the full field theory potential $\mathcal{U}(\xi)$ from the linearized potential $Q(r)$.

For the fitting procedure, we
first set the pion mass and pion decay constant to their physical values, $m_\pi = 138$ MeV and $f_{\pi}=186$ MeV. 
The model has two more independent parameters $e$ and $\epsilon$ which we use as independent variables. $e$ sets the energy and length scales
\begin{equation}
 \ell=\frac{2\hbar}{f_{\pi}e}.
\end{equation} 

Our procedure of finding the appropriate model is the following.
We choose the parameters describing the model ($e, \epsilon$) and the linearized potential $\{Q_n,L_n\}$ as independent variables which we vary 
to obtain the most physical model.

\begin{enumerate}
\item The value of  $e$ sets all scales needed for the asymptotic value of the potential 
\begin{equation}
 Q(r\to\infty)\to m^2=\frac{2m_{\pi}}{f_{\pi}e}.
\end{equation}
\item From a set of potentials steps $\{Q_n,L_n\}$ we calculate the resonances and scale them to their physical values
\begin{equation}
 \omega_{i,\text{ph}}=\frac{f_{\pi}e}{2}\omega_i.
\end{equation} 
 Here we chose $N=3$ (two steps) in all examples except for example 4 (where $N=4$, leading to much slower convergence).
\item From $Q(r)$ we reconstruct the potential $\mathcal{U}$, find the isoscalar charge radius $R_c$ and the energy (mass) of the nucleon $M$.
\item We calculate a function which measures the weighted relative distances (deviations) from the desired values:
\begin{equation}
 G(e, \epsilon, \{Q_n,L_n\}) = w_M^2 \frac{(M-M_\text{ph})^2}{M_\text{ph}^2} + w_R^2 \frac{(R_c-R_{c,\text{ph}})^2}{R_{c,\text{ph}}^2} + \sum_{n=1}^3 w_{n,r}^2\frac{(\Omega_n-\Omega_{n,\text{ph}})^2}{\Omega_{n,\text{ph}}^2} + 
w_{n,i}^2\frac{(\Gamma_n-\Gamma_{n,\text{ph}})^2}{\Gamma_{n,\text{ph}}^2}.
\end{equation} 
For the desired (physical) values we choose $M_\text{ph} = 932$ MeV (one-fourth of the mass of the helium nucleus, because the helium does not receive spin/isospin contributions and, therefore, is suitable for a comparison with the classical Skyrmion mass), $R_{c,\text{ph}} = 0.72$ fm, and the Roper values $\Omega_1 = 508$ MeV, $\Omega_2 = 778$ MeV, $\Omega_3 = 958$ MeV, as well as $\Gamma_1 = \Gamma_2 = \Gamma_3 = 300$ MeV (taking into account the rather large error for $\Gamma_2$ and $\Gamma_3$). \\
Further,  the $w_k$ are the relative weights of each observable.

\item We vary the variables $(e, \epsilon, \{Q_n,L_n\})$ and repeat the steps 2-4 to minimize the above distance function.
\end{enumerate}

In principle, by choosing at least the same number of parameters $(e, \epsilon, \{Q_n,L_n\})$ as the known values ($M, R_c, \omega_i$) to which we 
want to match our model, we should be able to find a minimum giving $G=0$. 
In practice, we have found this a very difficult task. The function $G$ can have discontinuities and long and twisted valleys. 
Sometimes the whole structure of resonances can change dramatically when the height of the potential is changed even insignificantly.
Moreover, some values are more important (measured with higher precision in experiments) so it might be better sometimes to change the weights. 
For example, the widths of the resonances are known with a considerable error, therefore the weights could be lowered. 
Different choices of weights gave us different minimized values of $G$.

In Table \ref{tab:examples}  we show some of the values for the observables we found. In the different examples, both the initial values of the parameters $(e, \epsilon, \{Q_n,L_n\})$ and the weights $w_k$ where chosen slightly differently.

\begin{table}
 \caption{Values of observables of examples found from minimizing the function $G$, along with the relative deviations from the target values, together with the corresponding fit values of the parameters $e$ and $\epsilon$.} \label{tab:examples}
{\scriptsize
\begin{center}
\begin{tabular}{rrrrrrrrrrr}
\hline\hline
&\multicolumn{2}{c}{Example 1}&\multicolumn{2}{c}{Example 2}&\multicolumn{2}{c}{Example 3}&\multicolumn{2}{c}{Example 4}&\multicolumn{2}{c}{Example 
5}\\
\hline
$M$&\hspace*{.5cm}\texttt{918.65} & \texttt{(-1.43\%)}\hspace*{.5cm}&\hspace*{.5cm}\texttt{931.82} & 
\texttt{(-0.02\%)}\hspace*{.5cm}&\hspace*{.5cm}\texttt{930.47} & \texttt{(-0.16\%)}\hspace*{.5cm}&\hspace*{.5cm}\texttt{873.74} & 
\texttt{(-6.25\%)}\hspace*{.5cm}&\hspace*{.5cm}\texttt{925.00} & \texttt{(-0.75\%)}\hspace*{.5cm}\\
$R_c$&\hspace*{.5cm}\texttt{0.7239} & \texttt{(0.54\%)}\hspace*{.5cm}&\hspace*{.5cm}\texttt{0.7199} & 
\texttt{(-0.02\%)}\hspace*{.5cm}&\hspace*{.5cm}\texttt{0.7200} & \texttt{(-0.00\%)}\hspace*{.5cm}&\hspace*{.5cm}\texttt{0.7200} & 
\texttt{(-0.00\%)}\hspace*{.5cm}&\hspace*{.5cm}\texttt{0.7200} & \texttt{(-0.00\%)}\hspace*{.5cm}\\
$\Omega_1$&\hspace*{.5cm}\texttt{467.89} & \texttt{(-7.90\%)}\hspace*{.5cm}&\hspace*{.5cm}\texttt{505.00} & 
\texttt{(-0.59\%)}\hspace*{.5cm}&\hspace*{.5cm}\texttt{506.83} & \texttt{(-0.23\%)}\hspace*{.5cm}&\hspace*{.5cm}\texttt{465.37} & 
\texttt{(-8.39\%)}\hspace*{.5cm}&\hspace*{.5cm}\texttt{484.23} & \texttt{(-4.68\%)}\hspace*{.5cm}\\
$\Gamma_1$&\hspace*{.5cm}\texttt{301.25} & \texttt{(0.42\%)}\hspace*{.5cm}&\hspace*{.5cm}\texttt{292.06} & 
\texttt{(-2.65\%)}\hspace*{.5cm}&\hspace*{.5cm}\texttt{287.90} & \texttt{(-4.03\%)}\hspace*{.5cm}&\hspace*{.5cm}\texttt{281.52} & 
\texttt{(-6.16\%)}\hspace*{.5cm}&\hspace*{.5cm}\texttt{297.26} & \texttt{(-0.91\%)}\hspace*{.5cm}\\
$\Omega_2$&\hspace*{.5cm}\texttt{758.72} & \texttt{(-2.48\%)}\hspace*{.5cm}&\hspace*{.5cm}\texttt{832.37} & 
\texttt{(6.99\%)}\hspace*{.5cm}&\hspace*{.5cm}\texttt{827.26} & \texttt{(6.33\%)}\hspace*{.5cm}&\hspace*{.5cm}\texttt{757.35} & 
\texttt{(-2.65\%)}\hspace*{.5cm}&\hspace*{.5cm}\texttt{792.35} & \texttt{(1.84\%)}\hspace*{.5cm}\\
$\Gamma_2$&\hspace*{.5cm}\texttt{286.54} & \texttt{(-4.49\%)}\hspace*{.5cm}&\hspace*{.5cm}\texttt{293.98} & 
\texttt{(-2.01\%)}\hspace*{.5cm}&\hspace*{.5cm}\texttt{289.32} & \texttt{(-3.56\%)}\hspace*{.5cm}&\hspace*{.5cm}\texttt{282.26} & 
\texttt{(-5.91\%)}\hspace*{.5cm}&\hspace*{.5cm}\texttt{296.57} & \texttt{(-1.14\%)}\hspace*{.5cm}\\
$\Omega_3$&\hspace*{.5cm}\texttt{1008.39} & \texttt{(6.37\%)}\hspace*{.5cm}&\hspace*{.5cm}\texttt{1133.36} & 
\texttt{(19.55\%)}\hspace*{.5cm}&\hspace*{.5cm}\texttt{1123.78} & \texttt{(18.54\%)}\hspace*{.5cm}&\hspace*{.5cm}\texttt{1014.10} & 
\texttt{(6.97\%)}\hspace*{.5cm}&\hspace*{.5cm}\texttt{1070.27} & \texttt{(12.90\%)}\hspace*{.5cm}\\
$\Gamma_3$&\hspace*{.5cm}\texttt{313.12} & \texttt{(4.37\%)}\hspace*{.5cm}&\hspace*{.5cm}\texttt{328.08} & 
\texttt{(9.36\%)}\hspace*{.5cm}&\hspace*{.5cm}\texttt{325.01} & \texttt{(8.34\%)}\hspace*{.5cm}&\hspace*{.5cm}\texttt{310.74} & 
\texttt{(3.58\%)}\hspace*{.5cm}&\hspace*{.5cm}\texttt{327.21} & \texttt{(9.07\%)}\hspace*{.5cm}\\
\hline
$e$\hspace*{0.1cm}  & \multicolumn{2}{c}{\texttt{ 1.2872}} 
& \multicolumn{2}{c}{\texttt{1.2936} } & \multicolumn{2}{c}{\texttt{1.2923} } & \multicolumn{2}{c}{\texttt{1.3320} } & \multicolumn{2}{c}{\texttt{1.2950} }\\
$\epsilon$\hspace*{0.1cm} & \multicolumn{2}{c}{\hspace*{0.1cm} \texttt{0.01251} } & \multicolumn{2}{c}{ \texttt{0.09602} } & \multicolumn{2}{c}{ \texttt{0.08431} } & \multicolumn{2}{c}{ \texttt{0.10886} } & \multicolumn{2}{c}{\texttt{0.07609} } \\
\hline
\hline
\end{tabular}
\end{center}
}
\end{table}

\begin{figure}

\includegraphics[width=0.9\textwidth,angle=0]{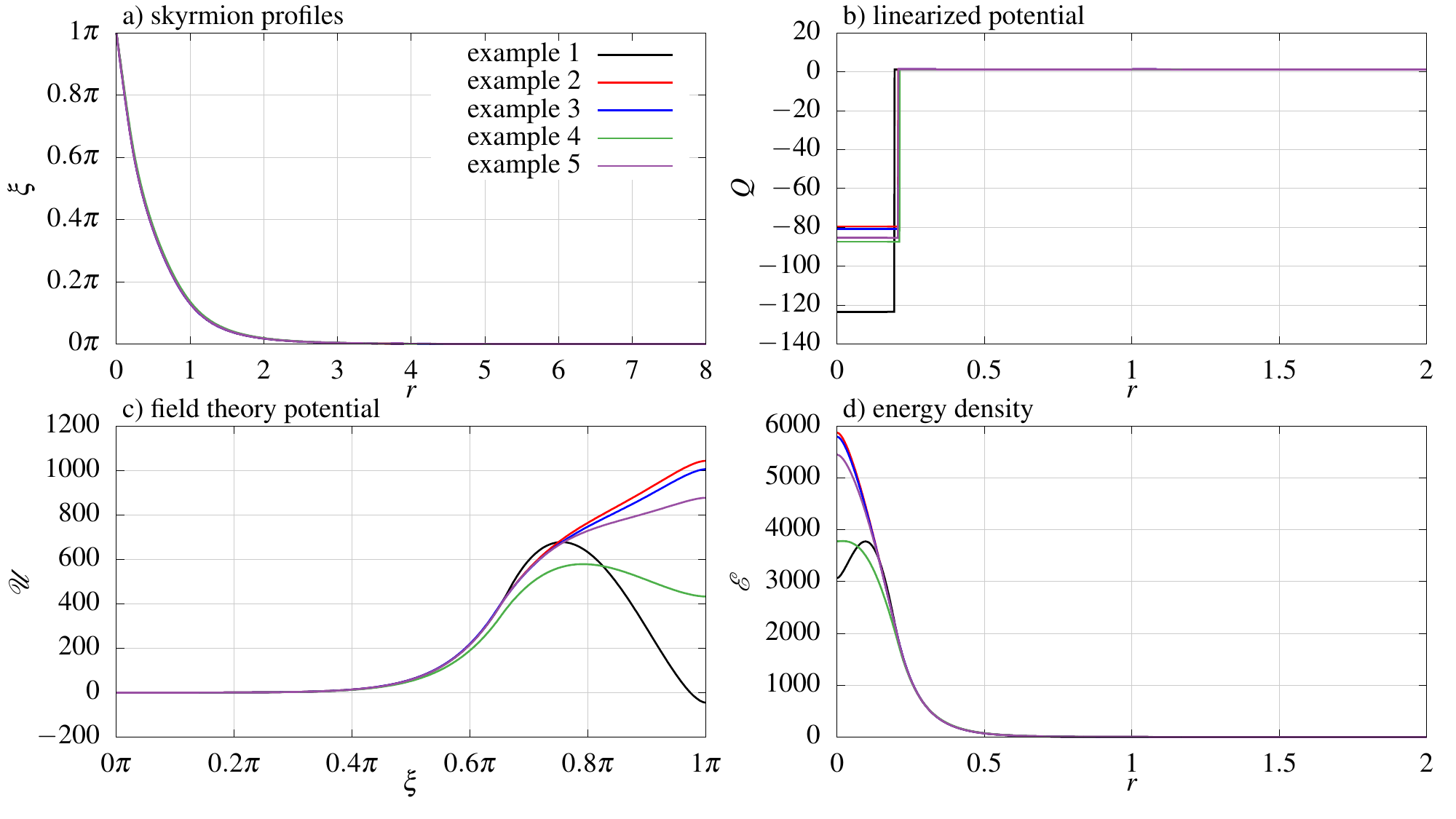}
\caption{\small Example potentials reconstructed from the given resonances. The plots are in Skyrme units.} \label{fig.reconstruction2}
\end{figure}

Figure \ref{fig.reconstruction2} shows the reconstructed skyrmion profiles $\xi(r)$, potential $Q(r)$, field theory potential $\mathcal{U}(\xi)$ and 
the energy density inside the skyrmion. Note that the profiles are almost indistinguishable.

\section{Summary}
In this paper, we investigated in detail the possibility to describe the Roper resonances within the context of the Skyrme model. We found that the quantitative results strongly depend on the particular Skyrme model under consideration and, specifically, on the potential term. For the standard pion mass potential, it turns out that the Roper resonances are not reproduced very well. In particular, the widths of the higher resonances always come out much larger than for the lowest resonance, in contrast to established experimental facts. From a theoretical point of view there is, however, no reason to restrict to the pion mass potential. The potential should always contain a mass term giving rise to the correct pion mass, but additional terms are perfectly legitimate. Once this possibility is considered, we found that already for rather simple potentials (e.g., the repulsive potential of Ref. \cite{Sp2}) the tension between calculated and experimentally measured Roper resonances can be slightly reduced.
However, none of the analysed potentials (with or without the sextic term) provided the correct Roper resonances. 

In a second line of investigation, we used an approach which is in some sense inverse to the one described above. Instead of departing from a given Skyrme model, we use a linear, quantum mechanical potential as a starting point. The determination of the Roper resonances from this linear potential is a much simpler problem. We then developed a reconstruction procedure which allows to reconstruct the Skyrme model (the field theoretical potential $\mathcal{U}$) from the initial linear potential $Q$. This second approach allows to describe the Roper resonances (and, in a second step, after the Skyrme model is reconstructed, also further observables) with a much higher precision, see Table \ref{tab:examples}. There are several directions in which this reconstruction method can and should be generalised. First of all, in this paper we only considered piecewise flat linear potentials $Q$, because this simplifies the calculation. Considering more general linear potentials should be interesting. Secondly, we were able to reconstruct the field theoretical potentials $\mathcal{U}$ only numerically. An analytical reconstruction, e.g., by expanding $\mathcal{U}$ in a certain set of basis functions, could be useful, because it would give us more analytical control over the resulting reconstructed Skyrme model.
We remark that the reconstruction method for the pure BPS Skyrme model (which had not been used in \cite{BPS-res} where only resonances for given potentials were calculated) might be useful to reduce the degeneracy in the potential parameter space and, thus, reconstruct physically viable potentials (although the BPS submodel by itself is insufficient as a low-energy effective field theory for strong interaction physics). 

As can be seen in Fig. \ref{fig.reconstruction2} (lower left panel), rather different Skyrme potentials can give rise to rather similar values for the physical observables (resonances, $M$, $R_c$). The observables considered in this paper are, therefore, in no way sufficient to pin down the correct physical Skyrme model potential or the coupling constants (relative strengths) of the derivative terms, and additional observables of hadron and nuclear physics are required for this task.
An interesting aspect of this problem is related to the relative strength of the sextic term (the value of the parameter $\epsilon$). Indeed, we argued in section V.B that higher values of $\epsilon$ should be expected, because they avoid very peaked energy densities. On the other hand, in the examples of fits to physical values of section V.C we always found rather small values of $\epsilon$ (see Table \ref{tab:examples}). The resolution of this apparent contradiction is easily understood. In the fit of section V.C we did not include observables which would impede peaked energy densities, and the energy densities resulting from the fit are  very peaked, see Fig.  \ref{fig.reconstruction2}. The simplest observable avoiding this energy concentration is the energy RMS radius $R_e$ defined by (here $E=\int d^3 x\, \mathcal{E}$ is the energy)
\be
R_e^2 =E^{-1} \int d^3 x \, r^2 \mathcal{E}  ,
\ee
because very peaked energy densities would lead to unacceptably small values for $R_e$.
We refrained from including this observable into the fit for two reasons. The first reason is that the inclusion of additional observables further complicates the numerical fit procedure. After all, the main purpose of the present article is a proof of principle, and not (yet) a detailed determination of the correct physical Skyrme model of nuclear and hadron physics. The second reason is that there does not seem to exist a direct measurement of $R_e$. The general expectation is that its value $R_{e,\text{ph}}$ should be slightly above $R_{c,\text{ph}}$, e.g., $0.8\, \text{fm} \le R_{e, \text{ph}} \le 1.0 \, \text{fm}$, but it is not clear which precise value to choose. Certainly, there exist further observables impeding very concentrated energy densities, because the physical expectation is that the energy density is rather flat in the core of a nucleus (nucleon), with a pion tail close to the surface.

Our main result is that generalised versions of the Skyrme model are perfectly capable of reproducing the Roper resonances and other observables with high precision. The further investigation of these generalised Skyrme models is, therefore, a promising and very timely research direction in the quest for a reliable low-energy effective field theory of strong interactions and nuclear physics.

\section*{Acknowledgements}
C.A. and A.W. acknowledge financial support from the Ministry of Education, Culture, and Sports, Spain (Grant No. FPA 2014-58-293-C2-1-P), the Xunta de Galicia (Grant No. INCITE09.296.035PR and Conselleria de Educacion), the Spanish Consolider-Ingenio 2010 Programme CPAN (CSD2007-00042), and FEDER. 
Some of the work of M.H. was undertaken at the Department of Mathematics
and Statistics, University of Massachusetts, financially supported by FP7,
Marie Curie Actions, People, International Research Staff Exchange Scheme
(IRSES-606096).


\begin{thebibliography}{20}
\bibitem{skyrme} T.H.R. Skyrme, Proc. Roy. Soc. Lon. {\bf 260},
127 (1961);  Nucl. Phys. {\bf 31}, 556 (1962); J. Math. Phys. {\bf
12}, 1735 (1971).
\bibitem{thooft} G. t'Hooft, Nucl. Phys. B{\bf 72}. 461 (1974); E. Witten, Nucl. Phys. B{\bf 160}, 57 (1979); E. Witten, Nucl. Phys. B{\bf 223}, 433 (1983).
\bibitem{bary} G.S. Adkins, C.R. Nappi, E. Witten, Nucl. Phys. B {\bf 228} (1983) 552; 
G.S. Adkins, C.R. Nappi, Nucl. Phys. B {\bf 233} (1984) 109.
\bb{more-baryon}
    A. Jackson, A.D. Jackson, A.S. Goldhaber, G.E. Brown, L.C. Castillejo
  Phys. Lett. B 154, 101 (1985);
    G. Holzwarth, B. Schwesinger
    Rep. Prog. Phys. 49, 825 (1986);
   I. Zahed, G.E. Brown
    Phys. Rep. 142, 1 (1986);
    B. Schwesinger, H. Weigel
    Nucl. Phys. A 465, 733 (1987);
H. Weigel, B. Schwesinger, G. Holzwarth,
Phys. Lett. B 168, 321 (1986);
B. Schwesinger, H. Weigel, G. Holzwarth, 
Phys. Rept. 173, 173 (1989).
\bibitem{light}
E. Braaten, L. Carson,
Phys. Rev. Lett. {\bf 56} (1986) 1897; Phys. Rev. D {\bf 38} (1988) 3525;
L. Carson,
 Phys. Rev. Lett. {\bf 66} (1991) 1406; 
T.S. Walhout,
 Nucl. Phys. A {\bf 531} (1991) 596;
O.V. Manko, N.S. Manton, S.W. Wood, Phys. Rev. C {\bf 76} (2007) 055203; 
R.A. Battye, N.S. Manton, P.M. Sutcliffe, 
S.W. Wood, Phys. Rev. C {\bf 80} (2009) 034323; C. Halcrow, Nucl.Phys. B {\bf 904} (2016) 106
\bibitem{lau} P.H.C. Lau, N.S. Manton, Phys. Rev. Lett. {\bf 113} (2014) 23; C.J. Halcrow, C. King, N.S. Manton,
Phys. Rev. C95 (2017) 031303.
\bibitem{bind}
C. Adam, C. Naya, J. Sanchez-Guillen, A. Wereszczynski,  Phys. Rev. Lett. {\bf 111} (2013) 232501; Phys. Rev. C {\bf 88} (2013) 054313.
\bibitem{Sp2} M. Gillard, D. Harland, M. Speight, Nucl.Phys. B {\bf 895} (2015) 272-287;
M. Gillard, D. Harland, E. Kirk, B. Maybee, M. Speight, Nucl. Phys. {\bf B917} (2017) 286. 
\bibitem{gud1} S. B. Gudnason, Phys. Rev. D {\bf 93} (2016) 065048;
S. B. Gudnason, B. Zhang, N. Ma, Phys. Rev. {\bf D94} (2016) 125004; S. B. Gudnason, M. Nitta, Phys. Rev. {\bf D94} (2016) 065018.
\bibitem{stars}
C. Adam, C. Naya, J. Sanchez-Guillen, R. Vazquez, A. Wereszczynski, {\em Phys. Lett. B} {\bf 742} (2015) 136;
C. Adam, C. Naya, J. Sanchez-Guillen, R. Vazquez, A. Wereszczynski,
 {\em Phys. Rev. C} {\bf 92} (2015) 025802.
 \bibitem{piette star} S. G. Nelmes, B. M. A. G. Piette, Phys. Rev. D {\bf 84} (2011) 085017; Phys. Rev. D {\bf 85} (2012) 123004.

\bb{PDG} K.A. Olive et al. (Particle Data Group), Chin. Phys. C38, 090001 (2014) (URL: http://pdg.lbl.gov)

\bibitem{pressure} C. Adam, M. Haberichter, A. Wereszczynski,  Phys. Rev. C {\bf 92} (2015) 055807.
\bibitem{term} 
C. Adam, C. Naya, J. Sanchez-Guillen, M. Speight, A. Wereszczynski,  Phys. Rev. D {\bf 90} (2014) 045003; 
C. Adam, T. Klahn, C. Naya, J. Sanchez-Guillen, R. Vazquez, A. Wereszczynski, Phys. Rev. D {\bf 91} (2015) 125030.
\bb{gold} A. Jackson, A.D. Jackson, A.S. Goldhaber, G.E. Brown, L.C. Castillejo, Phys. Lett. B {\bf 154} (1985) 101
\bibitem{BPS}
C. Adam, J. Sanchez-Guillen, A. Wereszczynski,
Phys. Lett. B{\bf 691}, 105 (2010); 
C. Adam, J. Sanchez-Guillen, A. Wereszczynski,
Phys. Rev. D{\bf 82}, 085015 (2010).

\bb{hajduk} C. Hajduk, B. Schwesinger, Phys. Lett. B 140 (1984) 172
\bb{kaulfuss} U. B. Kaulfuss, U.-G. Meissner, Phys. Lett. B154 (1985) 193
\bb{hayashi} A. Hayashi, G. Holzwarth, Phys. Lett. B140 (1984) 175
\bb{biedenharn} L. C. Biedenharn, Y. Dothan, M. Tarlini, Phys. Rev. D 31 (1985) 649
\bb{BPS vib}  C. Adam, C. Naya, J. Sanchez-Guillen, A. Wereszczynski, Phys.Lett. B726 (2013) 892
\bb{zahed} I. Zahed, U.-G. Meissner, U. B. Kaulfuss, Nucl. Phys. A 426 (1984) 525
\bb{nappi} J. D. Breit, C. R. Nappi, Phys. Rev. Lett. 53 1984) 889
\bb{piette} W.T. Lin, B. Piette, Phys.Rev. D77 (2008) 125028 [arXiv:0804.4786]
\bb{bizon} P. Bizon, T. Chmaj, A. Rostworowski, Phys.Rev. D75 (2007) 121702 [ math-ph/0701037]
\bb{BPS-res} 
C. Adam, M. Haberichter, T. Romanczukiewicz, A. Wereszczynski, 
Phys. Rev. D94 (2016) 096013. 

\bb{Q_Sk} M. Heusler, S. Droz, N. Straumann, Phys. Lett. B 271 (1991) 61.

\bb{Battye:2001} R.~A. Battye, P.~M. Sutcliffe, Rev. Math. Phys.,  {\bf 14} (2002) 29
\bb{Ascher:1981} U. Ascher,  J. Christiansen, R. Russell,  ACM TOMS, {\bf 7} (1981) 209; U. Ascher, J. Christiansen, R. Russell, Math. of Comp., {\bf 33} (1979) 639
\bb{SciPy} {\it ``Scipy, scientific tools for python, version 0.6.0''}, http://www.scipy.org

\bb{BPSvib2} T. Ioannidou, A. Lukacs, J. Math. Phys. {\bf 57} (2016)  022901.  
 \bb{DRS} P. Dorey, A. Halavanau, J. Mercer, T. Romanczukiewicz, Y. Shnir, 
JHEP 1705 (2017) 107.
 
\end{thebibliography}
\end{document}